\documentclass[twoside,twocolumn,9pt]{article}
\usepackage{extsizes}
\usepackage[super,sort&compress,comma]{natbib} 
\usepackage[version=3]{mhchem}
\usepackage[left=1.5cm, right=1.5cm, top=1.785cm, bottom=2.0cm]{geometry}
\usepackage{balance}
\usepackage{mathptmx}
\usepackage{sectsty}
\usepackage{graphicx} 
\usepackage{lastpage}
\usepackage[format=plain,justification=justified,singlelinecheck=false,font={stretch=1.125,small,sf},labelfont=bf,labelsep=space]{caption}
\usepackage{float}
\usepackage{fancyhdr}
\usepackage{fnpos}
\usepackage[english]{babel}
\addto{\captionsenglish}{%
  
}
\usepackage{textcomp}
\usepackage{times,mathptmx}
\usepackage{amssymb,amsmath}
\usepackage{gensymb}
\usepackage{array}
\usepackage{droidsans}
\usepackage{charter}
\usepackage[T1]{fontenc}
\usepackage[usenames,dvipsnames]{xcolor}
\usepackage{setspace}
\usepackage[compact]{titlesec}
\usepackage{hyperref}

\usepackage{epstopdf}

\definecolor{cream}{RGB}{222,217,201}

\begin{document}

\pagestyle{fancy}
\thispagestyle{plain}
\fancypagestyle{plain}{
\renewcommand{\headrulewidth}{0pt}
}

\makeFNbottom
\makeatletter
\renewcommand\LARGE{\@setfontsize\LARGE{15pt}{17}}
\renewcommand\Large{\@setfontsize\Large{12pt}{14}}
\renewcommand\large{\@setfontsize\large{10pt}{12}}
\renewcommand\footnotesize{\@setfontsize\footnotesize{7pt}{10}}
\makeatother

\renewcommand{\thefootnote}{\fnsymbol{footnote}}
\renewcommand\footnoterule{\vspace*{1pt}%
\color{cream}\hrule width 3.5in height 0.4pt \color{black}\vspace*{5pt}} 
\setcounter{secnumdepth}{5}

\makeatletter 
\renewcommand\@biblabel[1]{#1}            
\renewcommand\@makefntext[1]%
{\noindent\makebox[0pt][r]{\@thefnmark\,}#1}
\makeatother 
\renewcommand{\figurename}{\small{Fig.}~}
\sectionfont{\sffamily\Large}
\subsectionfont{\normalsize}
\subsubsectionfont{\bf}
\setstretch{1.125} 
\setlength{\skip\footins}{0.8cm}
\setlength{\footnotesep}{0.25cm}
\setlength{\jot}{10pt}
\titlespacing*{\section}{0pt}{4pt}{4pt}
\titlespacing*{\subsection}{0pt}{15pt}{1pt}

\fancyfoot{}
\fancyfoot[LO,RE]{\vspace{-7.1pt}\includegraphics[height=9pt]{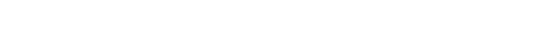}}
\fancyfoot[CO]{\vspace{-7.1pt}\hspace{13.2cm}\includegraphics{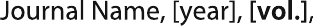}}
\fancyfoot[CE]{\vspace{-7.2pt}\hspace{-14.2cm}\includegraphics{head_foot/RF}}
\fancyfoot[RO]{\footnotesize{\sffamily{1--\pageref{LastPage} ~\textbar  \hspace{2pt}\thepage}}}
\fancyfoot[LE]{\footnotesize{\sffamily{\thepage~\textbar\hspace{3.45cm} 1--\pageref{LastPage}}}}
\fancyhead{}
\renewcommand{\headrulewidth}{0pt} 
\renewcommand{\footrulewidth}{0pt}
\setlength{\arrayrulewidth}{1pt}
\setlength{\columnsep}{6.5mm}
\setlength\bibsep{1pt}

\makeatletter 
\newlength{\figrulesep} 
\setlength{\figrulesep}{0.5\textfloatsep} 

\newcommand{\topfigrule}{\vspace*{-1pt}%
\noindent{\color{cream}\rule[-\figrulesep]{\columnwidth}{1.5pt}} }

\newcommand{\botfigrule}{\vspace*{-2pt}%
\noindent{\color{cream}\rule[\figrulesep]{\columnwidth}{1.5pt}} }

\newcommand{\dblfigrule}{\vspace*{-1pt}%
\noindent{\color{cream}\rule[-\figrulesep]{\textwidth}{1.5pt}} }

\makeatother

\twocolumn[
  \begin{@twocolumnfalse}
{\includegraphics[height=30pt]{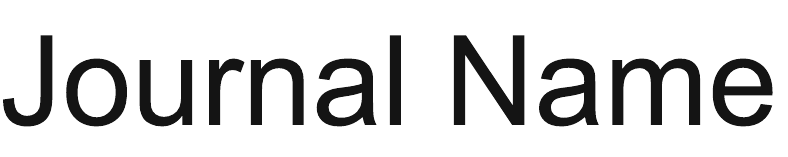}\hfill\raisebox{0pt}[0pt][0pt]{\includegraphics[height=55pt]{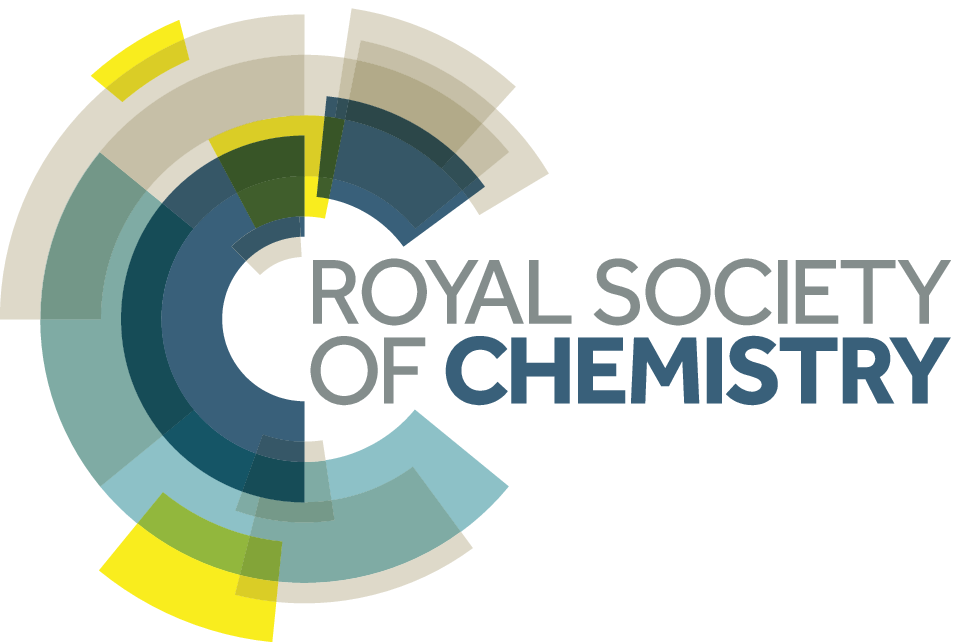}}\\[1ex]
\includegraphics[width=18.5cm]{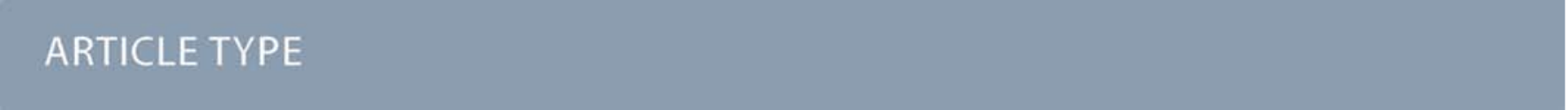}}\par
\vspace{1em}
\sffamily
\begin{tabular}{m{4.5cm} p{13.5cm} }

\includegraphics{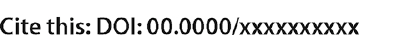} & \noindent\LARGE{{Assembly of nanocube super-structures directed by surface and magnetic interactions$^\dag$}} \\
\vspace{0.3cm} & \vspace{0.3cm} \\

 & \noindent\large{Igor Stankovi\'{c},\textit{$^{a,\ast}$} Luis Lizardi,\textit{$^{b}$} and Carlos Garc\'ia\textit{$^{b}$}} \\
 \\

\includegraphics{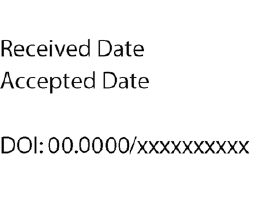} & \noindent\normalsize{We study the stabilisation of clusters and lattices of cuboidal particles with long-ranged magnetic dipolar and short-ranged surface interactions. Two realistic systems were considered: one with magnetisation orientated in the [001] crystallographic direction, and the other with magnetisation along the [111] direction. We have studied magnetic nanocubes clusters first in the limit of $T=0$~K intending to elucidate the structural genesis of low energy configurations and then analysed finite-temperature behaviour of the same systems in simulations. Our results demonstrate that dipolar coupling can stabilise nanoparticle assemblies with cubic, planar, and linear arrangements seen previously in experiments. While attractive surface energy supports the formation of super-cubes, the repulsion results in the elongated structures in the form of rods and chains. We observe the stabilisation of the ferromagnetic planar arrangements of the cubes standing on their corners and in contact over edges. We illustrate that minimal energy structures depend only on the size of the assembly and balance of surface repulsion and magnetic dipolar coupling. The presented results are scalable to different particle sizes and material parameters.} \\

\end{tabular}

 \end{@twocolumnfalse} \vspace{0.6cm}

  ]

\renewcommand*\rmdefault{bch}\normalfont\upshape
\rmfamily
\section*{}
\vspace{-1cm}


\footnotetext{\textit{$^{a}$~Institute of Physics Belgrade, University of Belgrade, Pregrevica 118, 11080 Belgrade, Serbia.}}
\footnotetext{\textit{$^{\ast}$ Email: igor.stankovic@ipb.ac.rs}}
\footnotetext{\textit{$^{b}$~Departamento de F\'isica \& Centro Cient\'ifico Tecnol\'ogico de Valpara\'iso-CCTVal, Universidad T\'ecnica Federico Santa Mar\'ia, Av. Espa\~na 1680, Casilla 110-V, Valpara\'iso,~Chile.}}

\footnotetext{\dag~Electronic Supplementary Information (ESI) available: [details of any supplementary information available should be included here]. See DOI: 00.0000/00000000.}

\footnotetext{\ddag~Additional footnotes to the title and authors can be included \textit{e.g.}\ `Present address:' or `These authors contributed equally to this work' as above using the symbols: \ddag, \textsection, and \P. Please place the appropriate symbol next to the author's name and include a \texttt{\textbackslash footnotetext} entry in the the correct place in the list.}


\section{Introduction}

Assembly of nanoparticles into target functional structures refers to the spontaneous formation of ordered patterns from disordered constituents. The self and directed assembly of magnetic particles carrying permanent dipolar moment are of great interest for many technical applications. In some cases, complex superstructures are created in a sequence of steps in which first particles are formed, and the interactions between nanoparticles are carefully tuned to steer whole self-assembly process to ultimately form macroscale ordered structure~\cite{fernandez2017three,overviewassembly1,small-nanodev,Shi_adfm.201902301}. The external field can drive assembly to various shapes using coils, conductive wires or meso- to microscale two-dimensional (2D) magnetised shapes with nanoparticles as building blocks~\cite{singh2014self, kulic,adfm201504749,Xue_adfm.201807658,Li.adfm.201903467}. Applications rely on outstanding assembly properties of magnetic nanoparticles, and therefore, the understanding of relevant energetic scales is crucial for designing processes including magnetorheological fluids~\cite{Butter2003}, high-density magnetic storage devices~\cite{MagneticRecording}, and tailored superlattices~\cite{small-array,Li2019acsnanolett}. The magnetic spheres and corresponding dipolar hard-sphere model, with a point-dipole at the centre of a spherically symmetric hardcore, is one the most studied systems both in the experiment and theoretically due to the simplicity of representation for particles with magnetic interactions. 

\begin{figure*}
\includegraphics[width=19cm]{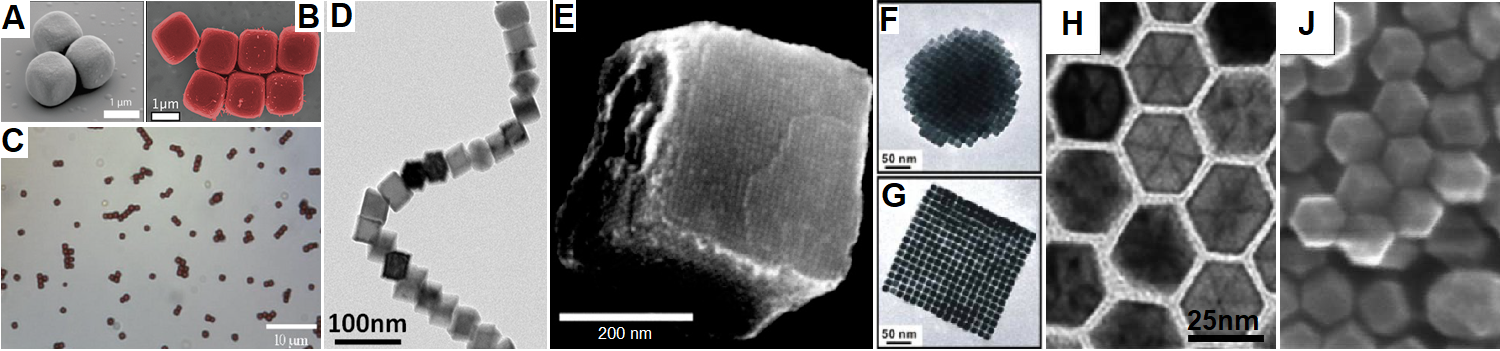}
\caption{SEM images of polymorphs composed of (A) silica-coated and (B) bare hematite microcuboids, adapted with permission from ref.~\cite{donaldson2018SM} published by The Royal Society of Chemistry and ref.\cite{Sacanna2012} copyright (2012) American Chemical Society, respectively. (C) Optical microscopy images of hematite cuboids during particle self-assembly induced by sedimentation, adapted with permission from ref.~\cite{Aoshima2012} copyright (2012) American Chemical Society. (D) The self-assembled chain in the gas phase of 25 nm iron/iron-oxide core/shell magnetic nanocubes, adapted with permission from ref.~\cite{NanoscaleBCL} published by The Royal Society of Chemistry. (E) SEM image of a 3D cuboid which consists of more than 10,000 iron-oxide nanocubes, adapted with permission from ref.~\cite{Mehdizadeh14484} copyright (2015) National Academy of Sciences. TEM images of self-organised (F) super-sphere along the [001] zone axis and (G) super-cube obtained by control of surface interactions, adapted with permission from ref.~\cite{Wang_ja308962w} copyright (2012) American Chemical Society. (H) Magnified TEM image and (J) STEM images of 2D nickel nanocube lattice (cubes are suspended on corners) both adapted with permission from ref.~\cite{Huaman2011} with permission from The Royal Society of Chemistry.
}  \label{fig:S0}
\end{figure*}

In contrast, cuboid particle geometry has received less attention despite advantageous for many applications in terms of photonic response~\cite{Li2019acsnanolett,Gao2012}, improved catalytic activity~\cite{catalyticcubes}, packing density~\cite{Disch2011,Ahniyaz17570}, and orientability~\cite{Mehdizadeh14484,singh2014self}. Still, the cubic shape is unique in two ways. First, this shape imposes strong coupling between geometry and magnetic interaction in assemblies of magnetic particles, where interparticle junctions formed by cube over their surfaces, corners, and edges are stabilised by strong attractive magnetic forces~\cite{Zhang2007, John2014, satoh2017modeling,donaldson2015directional}. Second, it is a compelling geometry for obtaining non-close-packed assemblies by designing surface interaction through facets. The magnetic forces are not screened in solution and are virtually independent of changes in experimental conditions such as humidity, pH, or solvent composition, which can alter surface interactions, thus, giving us significant design freedom\cite{Sacanna2012, Aoshima2012, donaldson2018SM}. The surface interactions between nanoparticles can be van der Waals, electrostatic or covalent interactions depending on composition of solvent and adsorbed layer on particles. A class of surface-modifying compounds that control the association of nanoparticles is referred to here as {\it ligands}, but also terms molecular linkers, surface modifiers and surfactants (i.e., surface active agents) are used elsewhere in literature~\cite{HEINZ20171,Heuer2019,Niculaes2017}. It is worth noting that particles with significant shape anisotropy can remain single domain at much larger sizes than their spherical counterparts~\cite{Bedanta2008}.
From a fundamental point of view, by looking into a self-assembled structure we can in-reverse conclude about interactions present in the system. There is an important reason to investigate mono-dispersed magnetic cubes: atomically flat sides of nanocubes allow them to glide almost without friction over their superstructures~\cite{NanoscaleBCL}. Besides, square symmetry reduces the number of local energetic minima in which the system can be quenched, preventing the creation of the clumps\cite{Mehdizadeh14484}, while tailoring the magnetic properties of the nanoparticles can provide an effective approach to direct self-assembly process\cite{Noh2012, Sacanna2012}. Here, we analyse magnetic nanocubes in which interaction landscape is defined by the steric effect, and magnetic and surface interactions\cite{NanoscaleBCL, Ahniyaz17570, Aoshima2012, Kostiainen2013, Luo_smll.201501783}.

{Figure~\ref{fig:S0}} summarises the magnetic nanocuboid structure stabilised by an interplay of density, magnetic anisotropy, the strength of dipolar coupling, and surface interactions. If the density of particles is low, small clusters might be created~\cite{Sacanna2012}. The hematite micron-sized cuboids demonstrate how dipolar interactions and particle shape result in the creation of the regular {\it polymorphs}, {cf.}, Figure~\ref{fig:S0}A-C. Designing a rational assembly mechanism based on magnetostatic interactions requires understanding differences between the energies of different structures. Strong and long-ranged dipolar coupling leads to the formation of macroscopic chains even in the gas phase, from particles coming from the cluster source, as demonstrated in Figure~\ref{fig:S0}D from Ref.~\cite{NanoscaleBCL}. The tendency for hydrophobic particles to cluster in water is readily used to self-assemble super-particles with remarkable internal order, {e.g.}, regular {\it super-cubes} composed of up to $10000$ nanocubes in Figure~\ref{fig:S0}E from Ref.~\cite{Mehdizadeh14484}. The shape of assemblies can be tuned between super-spheres and super-cubes, {cf.} Figure~\ref{fig:S0}F,G from Ref.~\cite{Wang_ja308962w}. Chemistry of solution in which particles are created, {i.e.}, presence of chloride ions and fatty acids, which control growth and prevent agglomeration result also in strong repulsion of the particles. Figure~\ref{fig:S0}H,J adapted from Ref.\cite{Huaman2011}, shows two-dimensional crystal formed by cubic-shaped particles standing on their corners, which is the result of the interplay of repulsion and magnetic interaction between particles.

In the present contribution, we consider assemblies of cubic magnetic particles and analyse the structural changes of minimal energy configuration. Designing a rational assembly mechanism based on magnetostatic interactions requires understanding differences between the energies of different structures. We systematically investigate clusters and lattices stabilised by magnetic dipolar coupling of their nanocube constituents. We compared the magnetic binding energy to the contact surface and calculated the magnitude of repulsive or attractive surface interaction needed to switch between different structures.

\section{Methods and Models}

We study a structure of the colloidal agglomerate formed from nano-particles described by Hamiltonian:
$H =  U + \epsilon_{\rm ss}S^{\rm int}+\gamma S^{\rm eff}+{\rm k}_{\rm B}T$, where $U$ is the dipolar magnetic interaction energy, $S^{\rm int}$ and $\epsilon_{\rm ss}S^{\rm int}$
are the interparticle surface contact area and surface binding energy, respectively. The effective surface area $S^{\rm eff}$, surface tension $\gamma$ of suspension liquid, and temperature $T$ are parameters which describe interaction with suspension.

\subsection{Dipolar magnetic interaction}

Magnetic cuboids are typically synthesised from iron and its oxides, as well as, from non-ferrous materials with different magnetic anisotropy and remanent magnetisation. An overview of sizes, saturation magnetisation, and easy magnetisation axis found in literature\cite{Sacanna2012,Ahniyaz17570,Song2004,MagneticRecording,abenojar2018magnetic,Aoshima2012,Huaman2011,Kronast2011,Noh2012,wu2016organic,Chen2004,chou2009controlled} is given in Table~\ref{table}. Magnetite (Fe$_{3}$O$_{4}$)~\cite{Gatel2015}, nickel platinum alloy~\cite{Huaman2011} nanocubes magnetic easy axis lying along the [111] crystallographic direction. This is in contrast to cubes of iron~\cite{Gatel2015}, FePt~\cite{Chen2004}, and cobalt/zinc ferrite~\cite{Noh2012,Song2004} that have cubic magnetic anisotropy and therefore preference for magnetisation along [001] direction. An additional variability of the properties may be achieved with core-shell structure, which can combine high remanent magnetisation of the core with magnetic anisotropy defined by the shell~\cite{Gatel2015}. The iron-oxide nanocuboids ({i.e.}, hematite, and magnetite) can be synthesised as micron-sized colloids. Hematite colloids, in particular, maintain a permanent dipole moment even at large particle size.

\begin{table}
\begin{tabular}{l@{\hskip1pt}c@{\hskip1pt}c@{\hskip1pt}c@{\hskip1pt}|l@{\hskip1pt}}
\hline\hline
nanocube&size& $M_s$& easy &reference \\
&[nm] &[kA/m]& axis & \\\hline\hline
FePt& 12 & 50 & [001] & Chou {\it et al.}\cite{chou2009controlled} \\\hline
CoFe$_2$O$_4$& 8 & 400 & [001]& Song and Zhang\cite{Song2004} \\
& 20 & 200 & & Wu {\it et al.}\cite{MagneticRecording} \\\hline
Zn$_{0.4}$Fe$_{2.6}$O$_4$& 20 & 874 & [001]& Noh {\it et al.}\cite{Noh2012}\\
& 60 & 1060 & &\\\hline
Fe&18&1700&[001]&Kronast {\it et al.}\cite{Kronast2011}\\\hline
NiPt & 25 & 600 & [111]& Cuya Huaman {\it et al.}\cite{Huaman2011}\\ \hline
Fe$_3$O$_4$ & 9 & 100 & [111] & Moya {\it et al.}\cite{Moya_2017} \\
magnetite & 14 & 130 & & \\
  & 30 & 160 & & Niculaes {\it et al.} \cite{Niculaes2017}\\
  & 1000 & 480 & & Aoshima {\it et al.}\cite{Aoshima2012}\\\hline
$\gamma$Fe$_2$O$_3$, maghemite&9&32&[111]&Ahniyaz {\it et al.}\cite{Ahniyaz17570} \\
$\alpha$Fe$_2$O$_3$, hematite&1000&2.2&[111]&Sacanna {\it et al.}\cite{Sacanna2012} \\
\hline\hline
\end{tabular}
\caption{Mean size, saturation magnetisation ($M_{\rm s}$), and easy magnetisation axis for different materials found in literature.}
\label{table}
\end{table}

Small single-domain magnets are treated like uniaxial magnets. We consider the two most common magnetisation directions: alongside [001] or along the principal diagonal [111] of the cube. Magnetic nanoparticles can have complex coupling involving both dipolar and exchange interactions. Their interaction is described through dipole-dipole interaction potential: it is assumed that each particle carries identical dipolar (magnetic) moment with magnitude $m_{\rm 0}=M_{\rm s}d^3$.  
\textcolor{black}{The saturation magnetisation is material and particle size-dependent and can take value from a modest 50~kA/m in 12~nm FePt~\cite{chou2009controlled} to 874~kA/m for 20 nm Zn$_{0.4}$Fe$_{2.6}$O$_4$ nanocubes~\cite{Noh2012}. We can approximate the reference magnetic energy of interaction of two touching nanocubes with 
$\upsilon=\mu_{\rm 0}m_{\rm 0}^2/4\pi d^3$. For dependence on saturation magnetization $M_s$  and $d$ particle size we obtain $\upsilon\propto M_{\rm s}^2d^3$. As a result, a choice of material or dimension of cubes has a strong influence on the magnetic interactions.}


In case of pure $40$~nm single-crystal magnetite cube and $M_{\rm s}=160$~kA/m, {cf.} Ref.~\cite{Niculaes2017}, the reference magnetic interaction energy was estimated to be $\upsilon=1$~eV, {i.e.}, 40k$_{\rm B}T$, where $T=300$~K is the temperature and k$_{\rm B}$ is the Boltzmann's constant. We chose the cube's dimensions to facilitate comparison both with real units used in experiment and scaled used in generic theoretical considerations. The magnetic field generated by one particle at the centre of mass of the other particle (placed side by side) is $B_{\rm 0}\approx\mu_{\rm 0}m_{\rm 0}/(2\pi d^3)=$16~mT. 

The dipolar magnetic interactions between an assembly of magnetic cubes are treated semi-analytically for small clusters $N\le8$ or using 9-dipole approximation~\cite{NanoscaleBCL} (see Supporting Information). This model represent an extension of the previously used single central dipole models\cite{John2014,Zhang2007}. In our model, the additional dipoles are placed in the corners of the cube and account for the interaction of the touching corners and edges of the cubes. The minimisation of energy was done systematically for $N\le16$ for all possible geometric and magnetic configurations. For larger systems, genetic algorithm was used to minimise magnetic configuration ({i.e.}, $N>16$). \textcolor{black}{We should note that in our calculations, we did not allow for the relaxation of magnetic moments around the easy-magnetisation axis. Such relaxation will be particularly pronounced in [111]-magnetic configurations where magnetisation has very localised flux closure, cf. Ref.~\cite{Mehdizadeh14484,Hakonsen_adfm201904825}.}

\subsection{Surface interaction}

The nanocube assemblies show a clearly defined contact surface area. These surfaces can be engineered repulsive or attractive, either by adsorbed layers from solution~\cite{Luo_smll.201501783,Kostiainen2013} or by polymers grafted on it~\cite{Gao2012,Maeda2015,Hakonsen_adfm201904825}. The stacking of the cubes tends to reduce or increase the surface area depending on the nature of surface interaction. In our calculations, surface energy is proportional to the contact surface. We will first discuss strong van der Waals attraction of the clean surfaces. Then we will explain how the surfaces can be modified to obtain weak attractive or even repulsive forces, and show that resulting energy scales are similar to that of magnetic interaction.


Clean metallic or metal-oxide nanocubes interact with each other through van der Waals interactions characterised by interaction energy and distance. The interaction energy parameter can be calculated as\cite{ISRAELACHVILI2011253}, $\epsilon_{\rm ss}=-A_{\rm mm}\sigma_{\rm mm}^4\rho^2/4\pi^2$ where $\sigma_{\rm mm}=$~0.35~nm is the size parameter for iron atoms, $\rho=$~85~nm$^{-3}$ the density of iron atoms in the bcc lattice, and $A_{mm}=$\textcolor{black}{1.321}8~eV is the average Hamaker constant for metals\cite{Baskin2012}. We obtain a value of $\epsilon_{\rm ss}$= -6 eV/nm$^2$, which result in a surface binding energy of $e_{\rm ss}=\epsilon_{\rm ss}d^2=$ 9.6~keV over a 40$\times$40~nm surface (fully touching cubes) and is comparable to the values found elsewhere~\cite{Gao2012, Maeda2015}. \textcolor{black}{We assume the resulting interaction is proportional to the contact surface and dependent on the orientation of the touching cubes\footnote{\citeauthor{Maeda2015}~\cite{Maeda2015} showed that, when cubes get close, dependence on the relative orientation intensify and interacting regions are sharply localised in the vicinity of the surface}.}

Since van der Waals interaction is strong, so-called steric stabilisation is used to control the coalescence of the particles typically by a thin adsorbed or grafted layer of appropriate thickness. These thin layers are used to obtain weak attraction, comparable to magnetic interaction energy between particles, and even repulsion, which can accommodate the formation of non-close-packed agglomerates. Engineering the repulsion and distance between magnetic particles is a typical way to steer the extent of the cube's aggregation\cite{NanoscaleBCL,Huaman2011}. The repulsive forces can arise from neutral steric layers ({i.e.}, entropic repulsion of grafted polymer chains) or electric double layers. The excluded volume of steric layers covering two particles results in a repulsive force. Each molecule of steric layer occupies a certain amount of space and, if molecules are brought close together, there is an associated cost in energy. The thickness of layers and the local density of grafted/attached polymers determines the extent and strength of repulsion, respectively. The maximal energy of the repulsion is of the order of $\epsilon_{\rm ss}=3$~meV/nm$^2$, {i.e.}, $e_{\rm s}=4.8$~eV for $40$~nm particles, {cf.} Ref.~\cite{Hakonsen_adfm201904825,Gao2012} We should also note that while the interaction between surfaces is repulsive, the grafted-polymer chains can accommodate edge contact. Gao {\it et al.}\cite{Gao2012} find weak and extremely short-ranged attractions between edges, which we did not include in our model. These edge-edge and edge-surface interactions will certainly further stabilise open structures analysed in the present work. We limited our considerations to a generic model for surface interactions which does not include any specific characteristics of many possible surface modifications~\cite{Gao2012,HEINZ20171,Heuer2019,Niculaes2017}.

\textcolor{black}{At this point, we would like to discuss how the size of the particles influences the balance between surface and magnetic energy. Several studies explored the assembly of the small nanocubes, i.e., $d\le$25~nm, cf. Ref.~\cite{singh2014self, NanoscaleBCL}. We recall that magnetic dipole interactions are roughly proportional to volume, {\i.e.}, $d^3$ and the surface interactions are proportional to $d^2$. In consequence, surface interactions would become more pronounced than magnetic interactions with decreasing particle size. Besides, for nanoparticles covered by a layer of organic ligands, the finite length of the ligand molecules become more critical in the assembly process while decreasing the particle size to ~20 nm and below. The interaction is repulsive with a distance of about $2-5$~nm for the grafting density of 0.04 chains/nm$^2$.\cite{Hakonsen_adfm201904825}. The effects of finite-range of surface repulsion are discussed in Sec. 3.5.}

\subsection{Interactions of assembly with suspension}

The interaction of hydrophobic particles with the suspension can itself alone drive self-assembly. Due to surface tension, a liquid tries to reduce the surface area at the interface. This also applies to the interaction of the suspension with the nano-agglomerate through their interface. The agglomeration process reduces interface surface between nanoparticles and suspension - synchronously reducing interparticle energy and surface energy of the liquid-nanoparticle interface. Water, the dominant chemical component of most suspensions, has particularly high surface tension. The Lum-Chandler-Weeks\cite{lum1999, Chandler2005} (LCW) theory explains the mechanism of how the interface surface area is reduced. Due to the thermal energy of molecules, according to LCW theory at ambient conditions (room temperature and $1$ atm pressure), liquid and vapour phases of water are close to phase coexistence. Therefore LCW theory bridges macroscopic wetting phenomena characterised by surface energy $\gamma S^{\rm eff}$ and nanoscale structuring of the suspension-nanoparticle interface, which gives rise to {\it effective surface area} $S^{\rm eff}$. The effective surface area $S_{\rm eff}$ is smaller than the total surface $S$ of the agglomerate. The factor of proportionality $\gamma$ is a solvent dependent parameter called surface tension. The direction in which self-assembly is going to drive the system is therefore also defined by effective surface area and surface tension. We should note that while the interaction between surfaces of individual particles (introduced in the previous section) is anisotropic and can be both attractive and repulsive, the interaction with solvent takes the form of isotropic external pressure. 

\subsection{Finite temperature simulations}

Molecular dynamics simulations in the LAMMPS simulation package were used to test the thermodynamic stability of investigated magnetic structures. We have performed three-dimensional simulations using a Langevin thermostat to keep constant-temperature conditions and include the effect of a fluid environment. The Langevin thermostat accounts both for viscous drag and for the random Brownian force on suspended particles exerted by surrounding fluid\footnote{\textcolor{black}{
We have based our model in the low-density limit, i.e., 
$4\pi k_{\rm B} T/\mu_{\rm 0}M_{\rm 0}^2\gg N/V$ where $N$ is the number of suspended particles in a volume $V$. A higher density of nanocubes will have a profound influence on the assembly. The long-ranged dipolar interaction can result in self-organized assemblies interacting with each other. There are two strategies to obtain assembly control: work with low concentration or reduce the volume in which particles are suspended in the moment of the assembly. The assembly at lower concentrations has a disadvantage that it will increase the assembly time since it would be governed by diffusion. The self-assembly in a reduced volume could be achieved by creating a flow of magnetic particles through a narrow tube or slit, either in liquid micro-reactor or a cluster gun, respectively. This will result in the creation of linear assemblies~\cite{NanoscaleBCL}. Another way to reduce volume is the assembly performed at liquid-gas or solid-liquid 2D interface ~\cite{Huaman2011}.}}. 

The cubes are represented by two types of contact potential: $(i)$ purely repulsive Weeks-Chandler-Anderson (WCA) potential, so called, contact potential~\cite{donaldson2015directional} and $(ii)$ repulsive or attractive potential in Yukawa form (see Supporting Information). The total force of contact due to the WCA potential is calculated using $33$ spheres (overlapping): a large central sphere with diameter $d$ and $32$ smaller with diameter $(\sqrt{3}-1)d/(\sqrt{3}+1)$ places in cube's corners and edges. This geometry allows smooth gliding of one cube over the other. The previous form of surface interactions, solely dependent on contact surface, is very convenient for analytic calculations but not straight forward to implement in molecular dynamics. Instead, we implement an additional interaction potential in Yukawa form between nine-point dipoles accounting for the short-range repulsion or attraction. This model was the basis for finite-temperature molecular dynamics simulation. The total force was the conservative force of inter-particle interactions. The magnetic interaction was treated with an interaction cut-off at $r_{cut}/d=8$, in order to reduce computational load without compromising precision {cf.} Ref.~\cite{satoh2017modeling}. We chose not to use the computationally expensive Ewald summation method, as our study is focused on the stability of assembled objects. The rotational degrees of freedom are also governed by the equations of motion for torque and angular velocity of the spheres. The total force and torque on each cube are computed as the sum of the forces and torques on its constituent particles at each timestep. The dipole orientation is accordingly rotated with the cube as a single entity. The rotation was implemented by creating internal data structures for each rigid body and performing time integration on these data structures~\cite{plimpton1995fast, Kamberaj2005}. 

The magnetic energy varies strongly with the system size, while for $40$~nm magnetite cube the reference magnetic energy is $\upsilon=1$eV ($M_{\rm s}=160$~kA/m, see Sec. 2.1), for $10$~nm magnetic cube is only $\upsilon=16$~meV. For this reason, we will make our discussion in this section independent of the system by introducing reference temperature $T_{\rm ref}=\upsilon/{\rm k}_{\rm B}$. The reference temperatures are $12000$~K and $1500$~K, for $40$~nm and $20$~nm cubes, respectively. It is worth to notice that the increase of temperature also may reduce saturation magnetisation $M_s$ which was not taken into account. The mass of the cuboid corresponded to $40$~nm magnetite cube and was distributed over nine constitutive dipolar particles. Molecular dynamics step was $t=20$~ms and the typical length of the molecular dynamics simulation $1000$~s. 

\begin{figure}
\includegraphics[width=8cm]{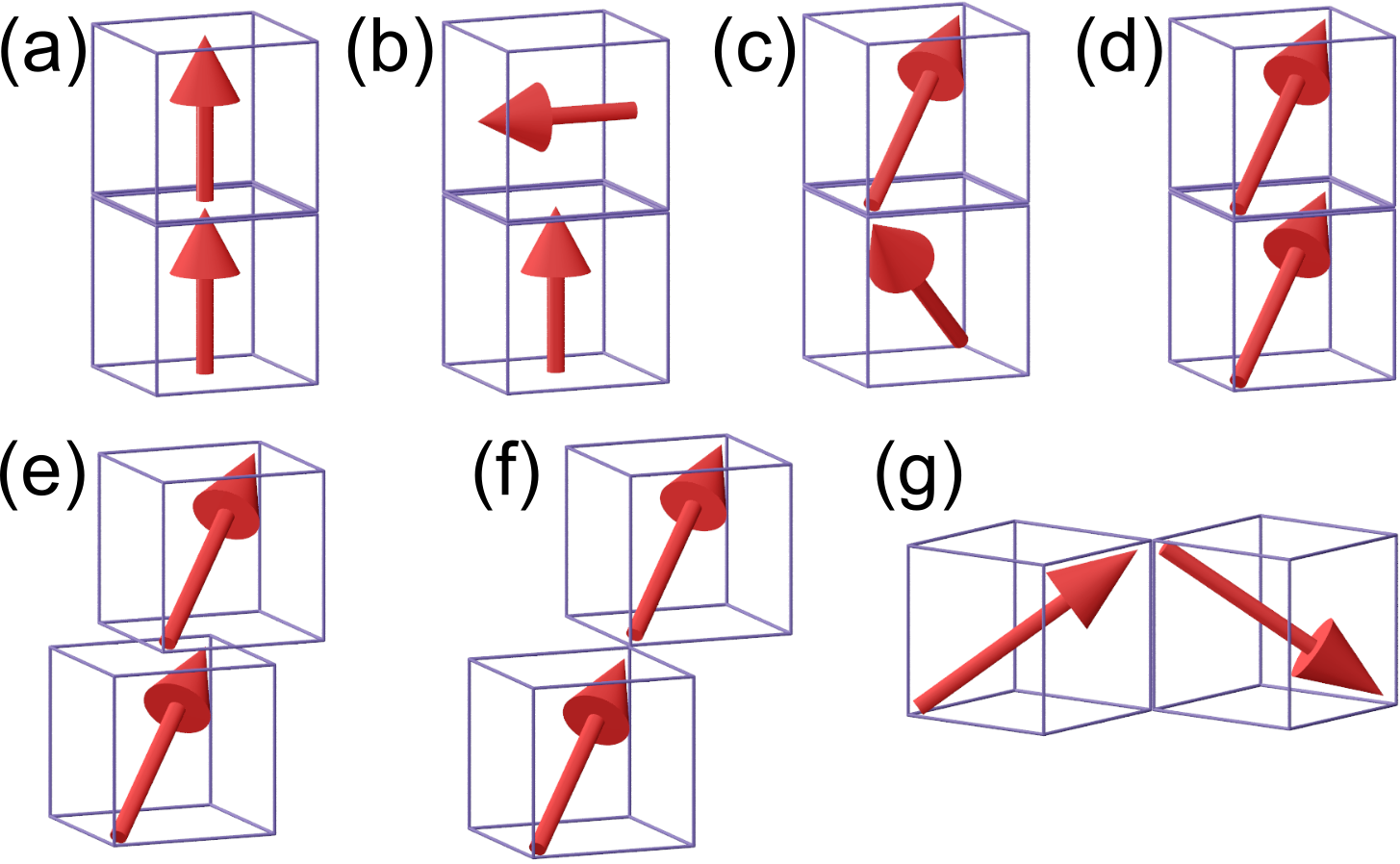}
  \caption{Different configurations of two uniformly [001] and [111] magnetised cubes. For [001] magnetisation, so called, (a) head-tail and (b) T-configuration are shown. In case of [111] magnetisation, so called, (c) zig-zag, (d) parallel, (e) head-tail, (f) point and (g) edge configurations are shown.}
  \label{fig:1}
\end{figure}

\section{Results}

The energy scales of assemblies can be probed in detail by analytical calculations of ideal configurations. We show system size and surface interaction dependent behaviour of different configurations for [111] and [001] magnetisations. Finally, we describe the behaviour of these systems when the surface tension effect and the dynamics in the solvent are taken into account.  

\begin{figure}
\includegraphics[width=8cm]{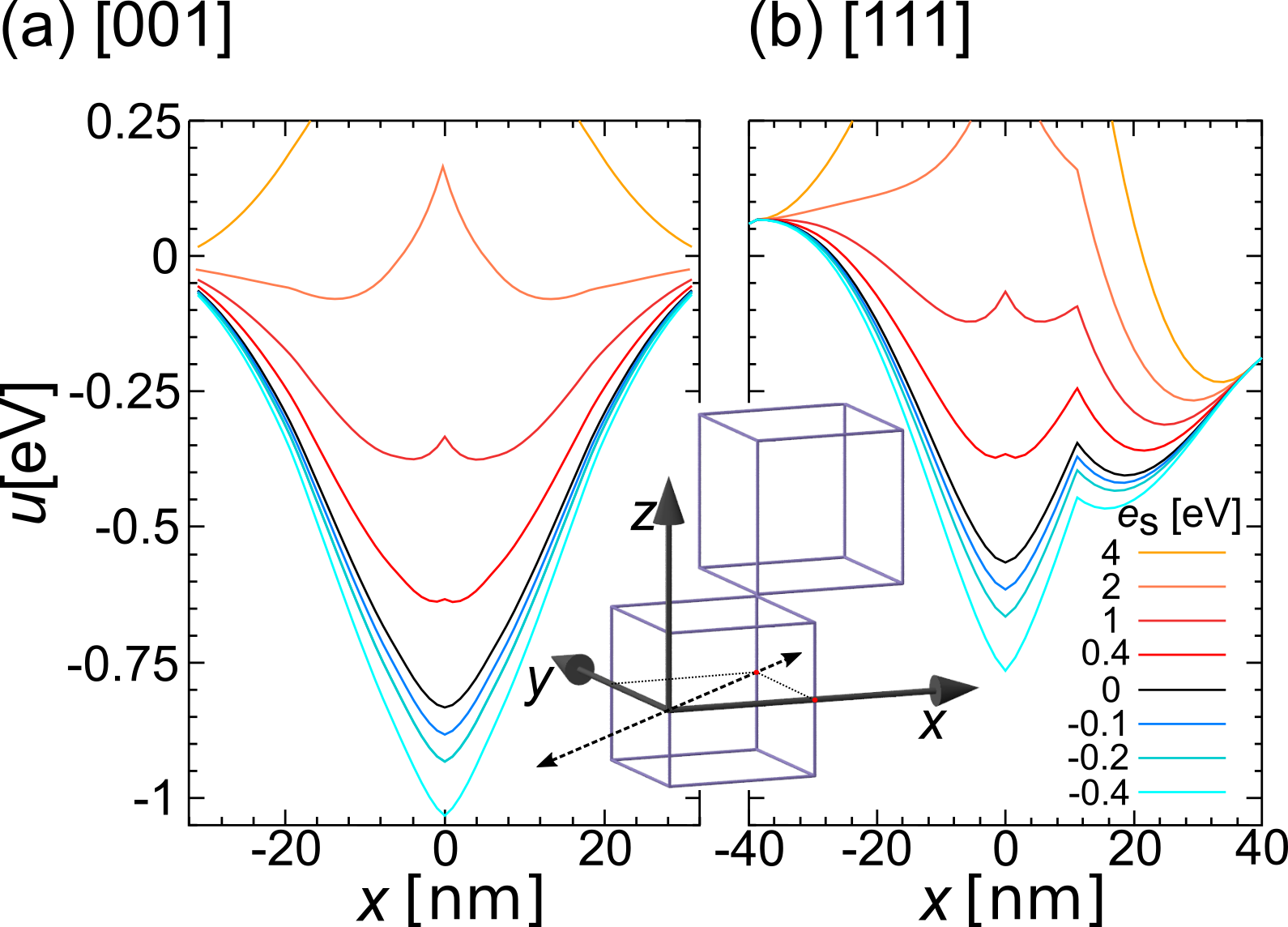}
  \caption{The minimal energy per particle $u$ for a cube moving along the face diagonal of the other one. We obtained that two cubes are always parallel and only magnetisation changes. The horizontal axis shows a projection of the centre of mass on the $x$-axis, total displacement is $\sqrt{2}x$. The magnetisation is oriented in the  (a) [001] and (b) [111] direction. The uniformly magnetised cubes with $d=40$~nm side are considered with $M_{\rm s}=160$~kA/m magnetisation. The energies are given in electron-volts, {i.e.}, the reference magnetic energy is $\upsilon=1$~eV. The surface coupling energy of two cubes in full contact $e_{\rm s}=\epsilon_{\rm ss}d^2$ is varied. The results are shown without $e_{\rm s}=0$, as well as for attractive $e_{\rm s}= $ -0.1, -0.2, -0.4~eV and repulsive $e_{\rm s}= $ 0.4, 1,2, 4~eV surface couplings.}
  \label{fig:2}
\end{figure}

\subsection{Energy of dimer}

It is instructive to first consider the interaction between two cubes with completely touching faces and compare it with the two elementary dipole situation. A ground configuration of the system is {\it head-tail} with magnetic energy $u^{\rm [001]}_{\rm min}=-0.815$~eV, see {Figure~\ref{fig:1}}(a). Opposite to head-tail configuration would be head-on-head configuration with the positive energy value $u^{\rm [001]}=0.815$~eV indicating that there is a strong energy penalty upon assembling a magnetic dimer in that configuration from infinite relative separation. Due to symmetry, the magnetic {\it T-configuration} in Figure~\ref{fig:1}(b) has zero energy. This feature is fully consistent with two spherical magnetic bids where magnetic T-configuration also has zero energy. The magnetic energy for the magnetic cube magnetised along principal diagonal is $u^{\rm [111]}_{\rm zz}=2u^{\rm [001]}/3=-0.555$~eV for the {\it zig-zag} magnetic configuration and zero for the parallel magnetic configuration. Both configurations are shown in Figure~\ref{fig:1}(c) and (d), respectively.

For a dimer of nanocubes, the assembly mechanism drives the particles to adopt structures that create a head-tail configuration, very much like chains of magnetic beads. In the case of [001] direction, the head-tail configuration represents a deep central minimum of potential energy, cf., {Figure~\ref{fig:2}}(a), restricting the lateral movement of the magnetic particles, and consequently leading to a quite stiff configuration. On the contrary, when the magnetisation is along the principal axis, {i.e.}, [111] direction, the structure is more flexible since magnetisation is pointing to/away of cube's corners. The resulting minimal magnetic energy profile of [111] magnetic cube dimmer has two minima, {i.e.}, local and global see cf., Figure~\ref{fig:2}(b).
The configuration with minimal energy has a zig-zag dipole vector placement, {i.e.}, Figure~\ref{fig:1}(c). At the minimum energy point (zig-zag configuration), the distance between their centres $\Delta r_2$ is equal to the cube size $\Delta r_2=d=40$~nm and particles are fully in contact, cf., Figure~\ref{fig:1}(c). The system can extend to a full head-tail configuration in Figure~\ref{fig:1}(e) by synchronous displacement and rotation of magnetic a cube between two minima\cite{NanoscaleBCL}. Balcells {\it et al.}\cite{NanoscaleBCL} showed that relative displacement and rotation can take place along the bottom of the circular valley connecting two minima characterised by gradual energy increase.

If we include surface energy, the total energy is a result of the net magnetic orientations and the contact surface. Now we analyse the evolution of the energy profile along the cube's face diagonal connecting two minima. The results are shown for different, so-called, {\it surface coupling} energies of two cubes in full contact $e_{\rm s}=\epsilon_{\rm ss}d^2$. We use surface coupling energy rather than energy per surface area to facilitate comparison with magnetic energy. If the particles attract each-other, {i.e.}, for $e_{\rm s}<0$ promotes fully touching configuration of the cubes (maximal contact surface). Therefore, the {\it attractive} surface interaction leads to a deeper central minimum of the potential energy, cf. curves for $e_{\rm s}<0$ in Figure~\ref{fig:2}, adding stability to ground state configuration, and consequently to increased stiffness of the configuration to any kind of deformation. 

If the surface interaction is {\it repulsive}, {i.e.}, for positive $e_{\rm s}>0$, the energy minimum becomes shallower. In the repulsive regime, two magnetic configurations [001] and [111] start to behave strikingly differently. For $e_{\rm s}>0$, the central minimum of [001] disappears and evolves into a potential valley with local conical energy maximum in the middle, see Figure~\ref{fig:2}(a). With increasing repulsion, the maximum increases, the stable configuration moves towards the edge of [001] magnetic nanocube and becomes fully unstable for $e_{\rm s}\gtrapprox2$~eV, {i.e.}, particles will not stay in close contact. 

Evolution of structure of [111] magnetised dimer in repulsive surface energy regime (i.e., $e_{\rm s}>0$) is complex. Figure~\ref{fig:2}(b) shows the evolution of the energy profile with surface coupling energy. for cubes with [111] magnetisation. The {\it head-tail} configuration in Figure~\ref{fig:1}(c) corresponds to the second minimum of the magnetic energy at point $\Delta r=0.5d=20$~nm in Figure~\ref{fig:2}(b). The centre of mass of the top particle is above the corner of the bottom particle in this configuration. At the second minimum, the centres of mass of the particles are $22\%$ further apart than at the global minimum (as shown in Figure~\ref{fig:2}, $\Delta r_2/d=\sqrt{3/2}$, {i.e.}, $\Delta r_2\approx49$~nm for 40~nm particles). The magnetic head-tail configuration of dipoles has magnetic energy $u^{\rm [111]}_{\rm ht}=-0.81$~eV, which is $27\%$ energy increase compared to the global minimum. The critical surface repulsion energy is equal to the difference of magnetic energies of two minima divided with relative contact surface difference of the two configurations $\epsilon_{\rm S}^{\rm cirt}=2(u^{\rm [111]}_{\rm ht}-u^{\rm [111]}_{\rm zz})/0.75\approx0.4$~eV. Below $e_{\rm s}=0.4$~eV particles will tend to stay in zig-zag magnetic configuration, and above they will open up to the head-tail configuration. The [111] magnetic particles will stay in contact even for very large repulsion energies, {e.g.}, see $e_{\rm s}=4$~eV curve in Figure~\ref{fig:1}(b). Finally, we want to find out which is the lowest magnetic energy configuration of the two [111] magnetised cube system if there is no contact surface. Figs.~\ref{fig:1}(f) and (g) show two [111] magnetised cubes touching only in a point or over an edge. The fully extended head-tail magnetic configuration in Figure~\ref{fig:1}(f) has $-0.125$~eV magnetic binding energy. The cubes touching over the edge have zig-zag magnetic configuration and a slightly lower binding energy $-0.14$~eV, {cf.} Figure~\ref{fig:1}(f).

\begin{figure}
\includegraphics[width = 8.5 cm]{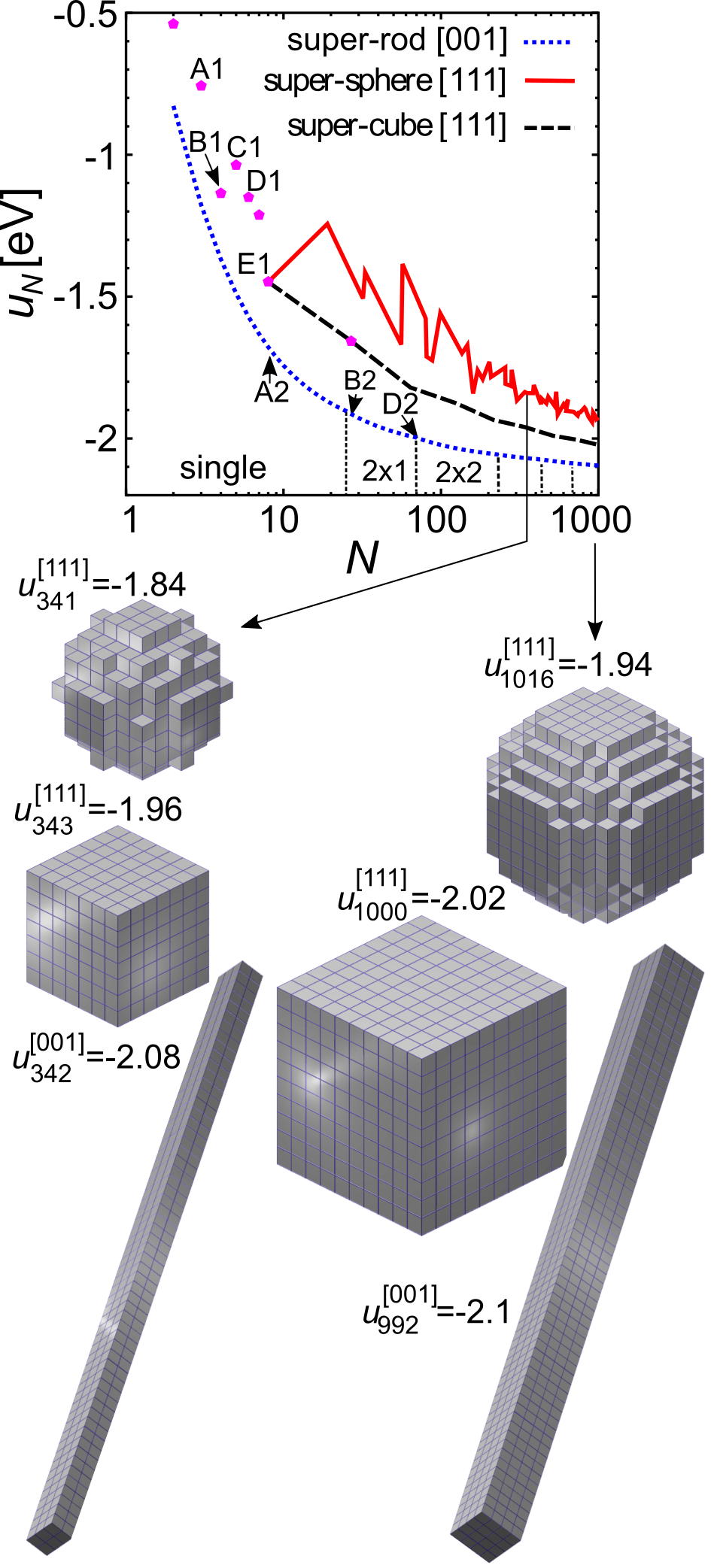}
  \caption{Magnetic binding energy per particle ($u_N$) as a function of number of cubes ($N$) in polymorphs (dots), super-cubes (dashed), super-spheres (full line) for magnetised cubes along the main diagonal ([111] direction) and minimal energy super-rods with [001] magnetised cubes. Regular isomers of super-structures are shown for $N\approx$\textcolor{black}{~343} and 1000 constituent nanocubes. The uniformly magnetised cubes with $d=40$~nm side are considered to have $M_{\rm s}=160$~kA/m magnetisation. The energies are given in electron-volts, {i.e.}, the reference energy is $\upsilon=1$~eV.}
  \label{fig:energy}
\end{figure} 

\begin{figure}
\includegraphics[width = 8.5 cm]{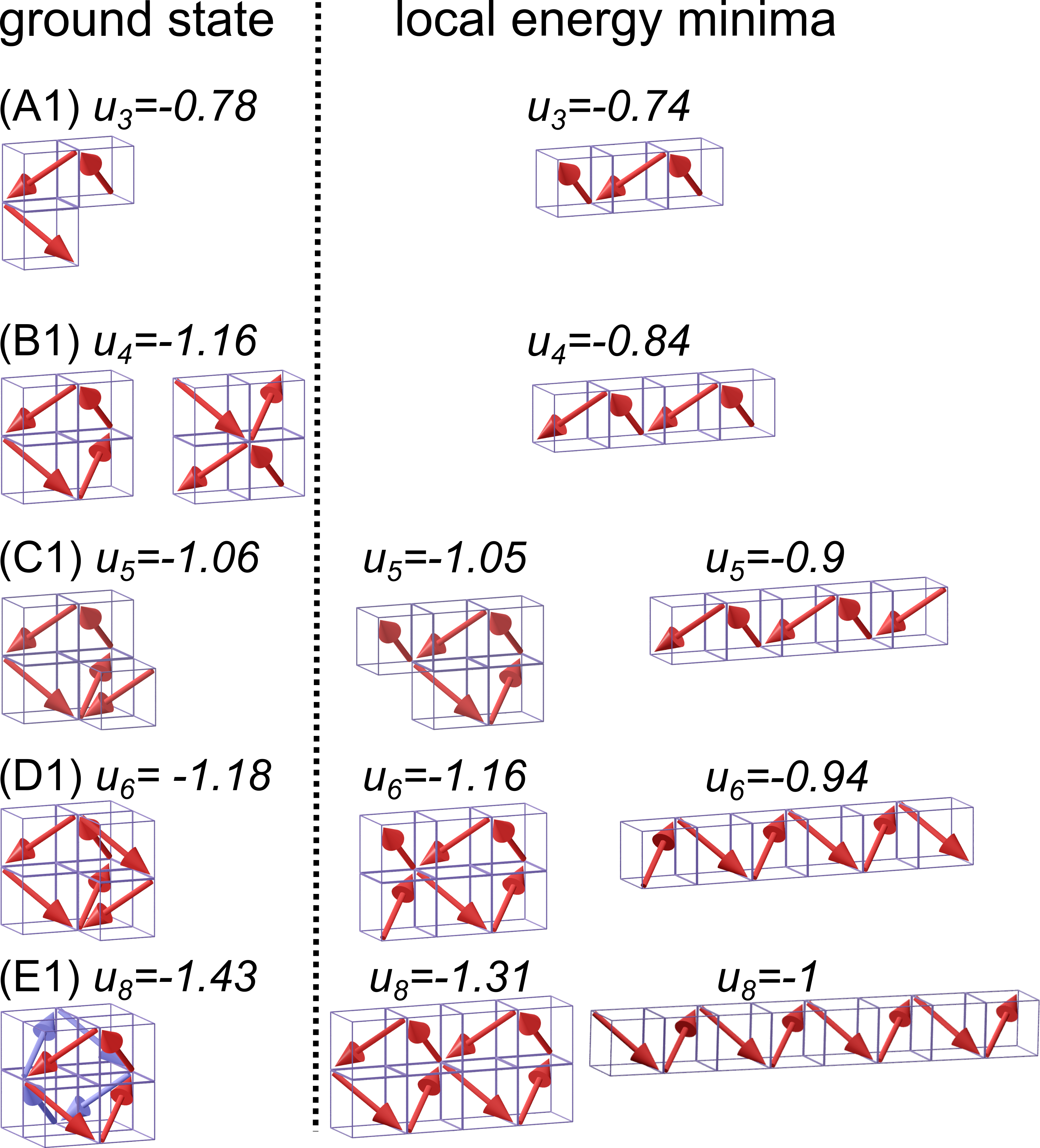}
  \caption{Magnetic binding energy per particle ($u_N$) of polymorphs with magnetisation in the [111] direction for $N= $3, 4, 5, 6, and 8. The ground state structures (left) and examples of non-ground state isomers (right) are shown. The uniformly magnetised cubes with $d=40$~nm side are considered with $M_{\rm s}=160$~kA/m magnetisation. The energies are given in electron-volts, {i.e.}, the reference energy is $\upsilon=1$~eV.}
  \label{fig:polymorps}
\end{figure}

\subsection{Magnetisation [111]}

\subsubsection{From polymorphs to super-cubes}

We have investigated ground state configurations of nanocube assemblies $2\le N\le8$ by calculating energies of all possible magnetic and geometrical arrangements. The diagonal polarisation of [111] magnetised cubes results in a strong tendency to create closed flux structures. The two-dimensional polymorphs were ground states only for $N=3,4$, {cf.} {Figure~\ref{fig:polymorps}}(A1) and (B1). For $N\ge5$, three-dimensional polymorphs become a ground state. For $N=5$, see Figure~\ref{fig:polymorps}(C1), the system has an uncompensated magnetic moment. Interestingly, the formation of five particle polymorph leads to penalty (increase) in magnetic binding energy, {i.e.}, binding energy per particle increases from $u_4=-1.16$~eV to $u_5=-1.06$~eV. Also we should note that there is less than $1\%$ binding energy difference for $N=5$ and $6$ polymorphs between 3D ({e.g.}, ground state) and 2D (non-ground state isomer) structures, {e.g.}, $u_5=-1.06$~eV for the cube on top of the $2\times2$ square and $u_5=-1.05$~eV on the side, see in Figure~\ref{fig:polymorps}(C1). Further closure of head-tail magnetic circulation of all magnetic cubes, {i.e.}, for $N=8$, leads to a $10\%$ difference in magnetic binding energy between 2D and 3D structures. More specifically, for polymorphs, a well-known arrangement of the dipoles in head-tail circulation is found (see Figure~\ref{fig:polymorps}). Direct computation of energies for all possible configurations increases exponentially with the number of cubes so, beyond eight nanocubes this approach is not feasible. We used the genetic algorithm~\cite{MESSINA201710} to obtained the minimal binding energy of a super-cube composed of $3\times3\times3$ nanocubes, see the SI. All three dipole vector components along sides of the cubes are inscribed antiferromagnetically ({cf.}, $N=8$ in Figure~\ref{fig:polymorps}), resulting in a vortex magnetic structure. For a super-cube composed of $3\times3\times3$ nanocubes, {e.g.}, $N=27$, the minimal binding energy was obtained by genetic algorithm~\cite{MESSINA201710} (see Supporting Information). All three dipole vector components along sides of the cubes are inscribed antiferromagnetically ({cf.}, $N=8$ in Figure~\ref{fig:polymorps}), resulting in a vortex magnetic structure. 

\subsubsection{Interparticle interactions: super-cube vs. super-sphere}

In {Figure~\ref{fig:energy}}, we have numerically compared the magnetic binding energy $u_N$ (per particle) of super-spheres and super-cubes. To calculate magnetic energy, the super-cubes and spheres were inscribed antiferromagnetically with [111] magnetisation. The energy evolution with the number of nanocubes is displayed in Figure~\ref{fig:energy}. We demonstrate that, with respect to magnetic binding, the super-sphere is an unfavourable configuration. The energy of a super-sphere is always above the energy of a super-cube (except in case of the smallest the sphere, {i.e.}, $N=8$, which is also a cube).

\begin{table}
\begin{tabular}{lclc}
\hline\hline
super structure & axis & ~~~~~$u_N$~[eV] & $S_N/d^2$ \\
\hline\hline
super-cube & [111] &$-\textcolor{black}{2.15}+\textcolor{black}{1.307}N^{-1/3}$ & $6N^{2/3}$\\
  & [001] &$-\textcolor{black}{2.15}+\textcolor{black}{1.321}N^{-1/3}$ & \\
\hline
super-sphere & [111] &$-\textcolor{black}{2.15}+\textcolor{black}{2.23}N^{-1/3}$&  $\approx7.4N^{2/3}$\\
& [001] & $-\textcolor{black}{2.15}+\textcolor{black}{2.3}N^{-1/3}$& \\
\hline
super-rod & [001] & $-\textcolor{black}{2.15}+\textcolor{black}{0.56}N^{-1/3}$ & $\approx\textcolor{black}{10.24} N^{2/3}$ \\
\hline
corner-cube & [111] &$-0.91~~+0.9N^{-1/3}$ & 0\\
\hline
zig-zag chain & [111] &$-1.17+1.3N^{-1}$&$5N+1$\\
head-tail chain & [001] & $-2.01+0.95N^{-1}$& \\
\hline
head-tail chain &[111] &$-0.95+1.2N^{-1}$&$(11N+1)/2$\\
\hline\hline
\end{tabular}
\caption{Scaling laws for magnetic energy per particle $u(N)$ and surface area $S(N)$ for different structures with the number of constitutive cubes ($N$). The results are shown for two easy magnetisation axes [111] and [001] in different structures such as; super-cube, super-sphere, super-rod (all three shown in Figure~\ref{fig:energy}), corner-cube planar structures [111] (Figure~\ref{fig:corner}), linear zig-zag [111] (right side of Figure~\ref{fig:polymorps}), head-tail [001] chains of fully touching cubes (e.g., in Figure~\ref{fig:rods}), and head-tail [111] chains of partially touching (over 1/4 of the side surface) cubes. The expressions for energies, surface areas of super-sphere and super-rod are obtained numerically from calculated data. The energies are given in electron-volts for uniformly magnetised cubes with $d=40$~nm side and saturation magnetisation $M_{\rm s}=160$~kA/m, {i.e.}, the reference energy is $\upsilon=1$~eV.}
\label{table1}
\end{table}

First, we discuss the situation when the surface coupling is attractive and much stronger than dipolar magnetic interactions between nanocubes. This is a limiting case where the system behaviour is independent of magnetic anisotropy (in our case, [111] or [001]). When surface coupling is attractive, {i.e.}, $e_{\rm s}<0$, the assemblies tend to decrease total (free) surface area. One can construct, so-called, {\it super-sphere} in which magnets are centred around the middle of the super-sphere within a radius $R$ ({cf.} Figure~\ref{fig:energy}). 

The perfect super-sphere formed from nanocubes has a surface consisting of kinks and steps. As a result, the surface area of the super-sphere is larger than that of a super-cube. The super-sphere surface ($S_{\rm ss}$) to volume ($V_{\rm ss}$) ratio is $S_{\rm ss}/V_{\rm ss}\approx7.4N^{-1/3}$, for $N\gg1$ and therefore about $23\%$ larger than that of the super-cube, $6N^{-1/3}$ made of an identical number of nanocube. This means that the creation of a super-cube (rather than a super-sphere) is a path to reduce free surface in a system composed of nanocubes. From scaling laws in {Table~\ref{table1}} the difference between the super-particles decreases with $N^{-1/3}$. Still, since the super-sphere has a larger exterior surface and hence a smaller contact surface than its cubic counterpart the surface repulsion needed to induce transformation is $e_{\rm s}>2.9$~eV and almost size independent. Therefore, the only way to obtain super-sphere is through interaction with the suspension, as we will show later.

\begin{figure}
\includegraphics[width = 6.5 cm]{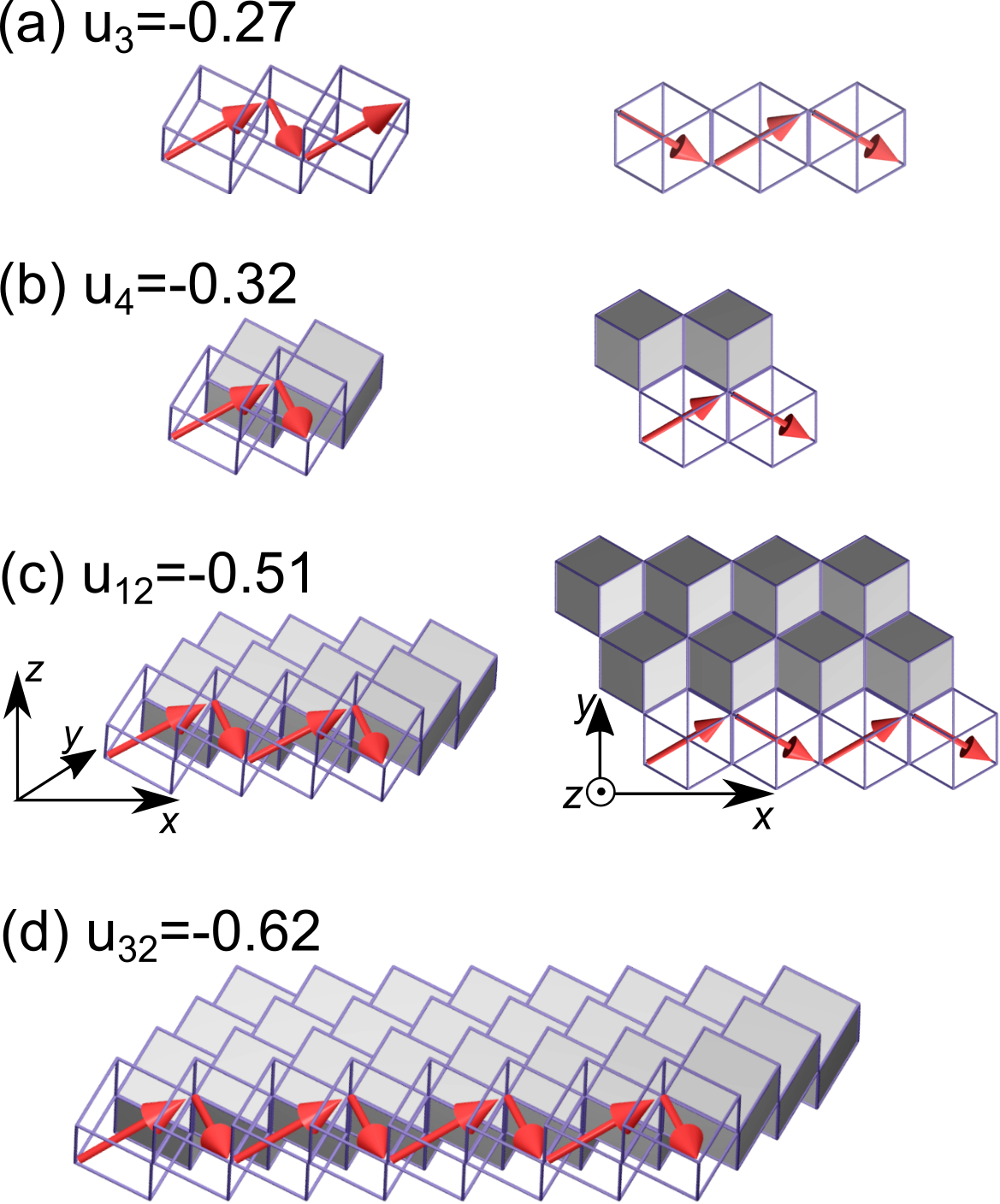}
\caption{Overview of the calculated magnetic energies per particle $u_N$ of corner-cube flakes with $N=$3, 4, 12, and 32 magnetic cubes. Side view (left) and top view (right) of the structures are given. One could observe a hexagon pattern seen in the experiment ({cf.}, Figure~\ref{fig:S0}K) in the top view. The uniformly magnetised cubes with $d=40$~nm side are considered to have saturation magnetisation $M_{\rm s}=160$~kA/m. The energies are given in electron-volts, {i.e.}, the reference energy is $\upsilon=1$~eV.}
  \label{fig:corner}
\end{figure}

\begin{figure}
\includegraphics[width = 8 cm]{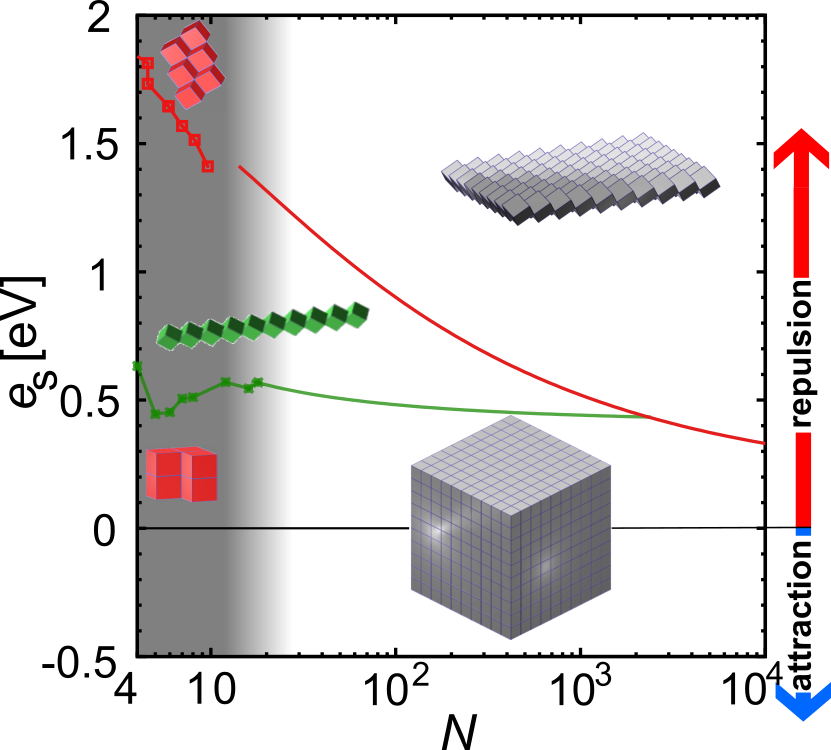}
  \caption{Phase diagram of magnetic cubes, with a surface coupling $e_{\rm s}=\epsilon_{\rm s}d^2$, plotted as a function of the cluster size. The (green) squares, connected as a guide to the eye, represent polymorphs being comprised only of a few particles located in the grey area which transcend directly into the chains of partially touching cubes with head-tail magnetic configurations. The (red) circles denote the upper limit of stability of head-tail chains above which they transform into corner-cube 2d structures. The right side of the diagram follows the phase transformation of large structures ($N\ge27$). The (green) full curve delimits area of stability of super-cubes and chains and (red) dashed curve same for chains and corner-cube 2d structures. The uniformly magnetised cubes with $d=40$~nm side are considered to have saturation magnetisation $M_{\rm s}=160$~kA/m.
  }
  \label{fig:diag111}
\end{figure}

\subsubsection{Phase behaviour of [111] magnetised nanocubes: cluster size effects}

Table~\ref{table1} lists energy scaling laws for different super-structures. It is found that the reduced energy per cube $u_N$ is given by $u_N=-\textcolor{black}{2.15}+\textcolor{black}{1.307}N^{-1/3}$ for super-cubes, which is slower than in previously studied systems of magnetic tubes~\cite{StankovicSM16,StankovicNanoscale}. The reduced energy per magnet is $u^{\rm 3D}_{\infty}=-\textcolor{black}{2.15}$~eV in a bulk material, $u^{zz}_{\infty}=$-1.17~eV in an infinite chain of the fully touching cubes (zig-zag magnetisation) and, $u^{\rm ht}_{\infty}=-0.95$~eV in head-tail chains. We found that the scaling laws for chain structures, $u_N\sim N^{-1}$, are in accordance with previous observations for one-dimensional dipolar structures made\cite{MESSINA201710}. 
 
To study the evolution of arrangements of [111] magnetised nanocubes with the strength of repulsive surface interaction we used scaling laws and calculated exact values of energy for different arrangements of polymorphs (i.e., $N<27$). We start from minimal magnetic energy structures described in the previous section and analyse the difference in energy resulting from the surface coupling between particles, {\it ie.}, $e_{\rm s}=\epsilon_{\rm ss}d^2$. When the surface repulsion increases $e_{\rm s}\gtrapprox0.6$, the magnetic assemblies tend to increase surface area.

The assemblies with $N<400$ will transcend into chains with partially touching cubes and head-tail magnetic configurations. This is somewhat surprising since the chains with fully touching faces have higher binding energy. Nevertheless, four times smaller contact surface of partially touching head-tail magnetised cubes makes them favourable as a minimal energy state over fully touching chains. The limiting energies are denoted with (green) squares in {Figure~\ref{fig:diag111}}. With increasing repulsive coupling we observe a transition to, so-called, {\it corner-cube} flakes. This remarkable transformation originates from the necessity to avoid any surface-to-surface contact while simultaneously increasing binding energy by minimising the distance between cubes. The transition from $4\times4$ square to corner-cube square takes place for a surface coupling energy of $e_{\rm s}=0.84$~eV. {Figure~\ref{fig:corner}} gives an overview of corner-cube flakes. The constitutive chains have the magnetisation parallel to the patch longer side and ferromagnetic configuration. The magnetic binding energies are considerably higher than in the case of polymorphs, $u^{cc}_{4}=-0.32$~eV, $u^{cc}_{12}=-0.32$~eV, and $u^{cc}_{32}=-0.62$~eV, see Figure~\ref{fig:corner}.

Our analysis shows that chains are energetically favoured transient structure only for small assemblies. For a large number of particles, {i.e.}, $N>400$, the chain vanishes as an intermediate structure and there is a direct transition to corner-cube flakes, {cf.} Figure~\ref{fig:diag111}. With an increasing number of magnetic cubes the binding energy of the corner-cube flake is converging to value of $u^{\rm 2D, cc}_{\infty}=-0.91$~eV. This binding energy is only about \textcolor{black}{42$\%$} of the value for bulk magnetic cube structure, {i.e.}, \textcolor{black}{$u^{\rm 3D}_{\infty}=-2.15$}~eV. The resulting limiting surface energy required to unrolling super-cube into a corner-cube surface converges to \textcolor{black}{$e^{\rm[111]}_{\rm cc}\approx(e^{\rm [111]}_{\rm corner}-e^{\rm 3D}_{\infty})/3=0.41$~eV}, {cf.} in Figure~\ref{fig:diag111}.

\subsection{Magnetisation [001]}

\subsubsection{From chain to rods}

\begin{figure}
\includegraphics[width = 8.5 cm]{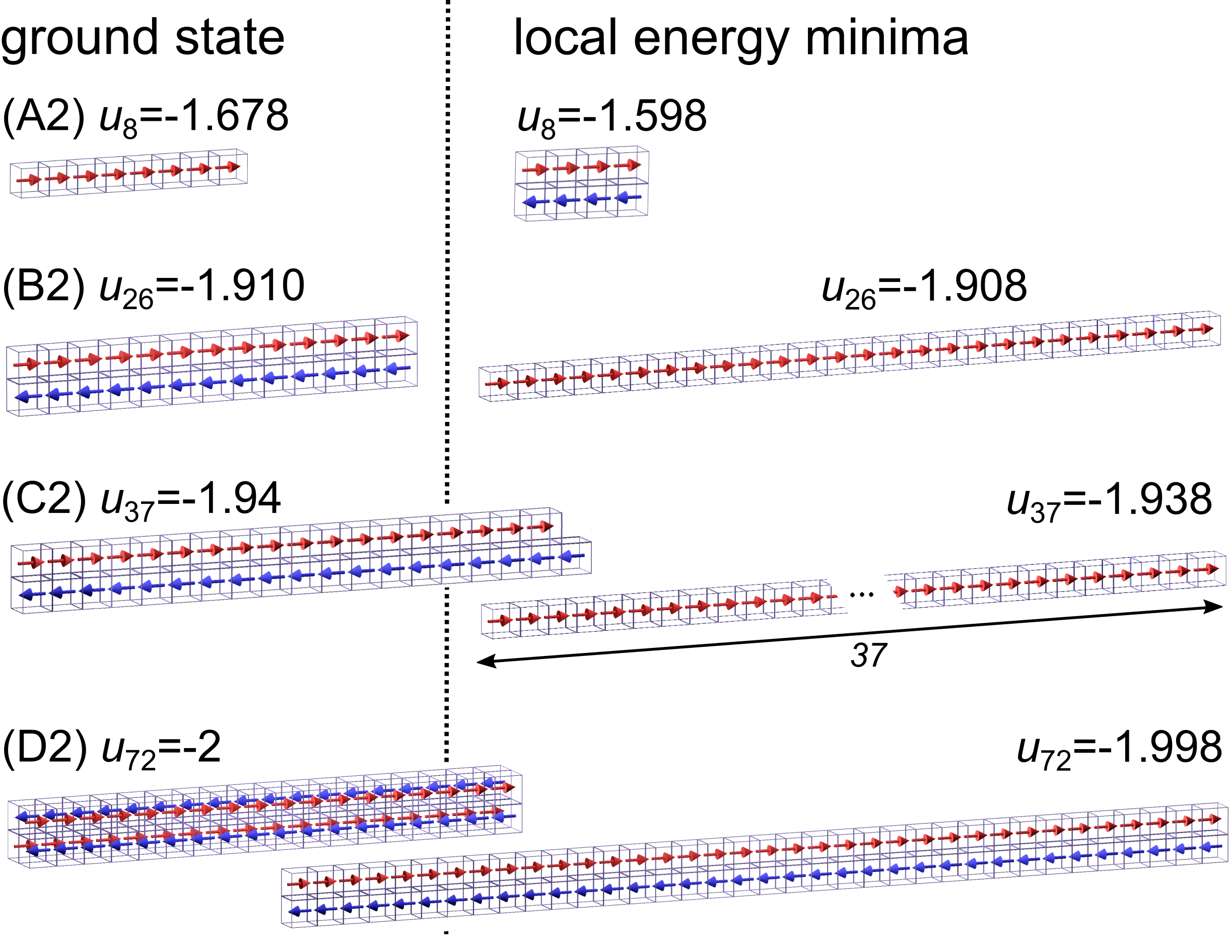}
  \caption{Magnetic energy per particle ($u_N$) as a function of number of cubes ($N=$ 8, \textcolor{black}{26, 37, and 72}) with magnetisation in the [001] direction. The ground state structures (left) and examples of non-ground state isomers (right) are shown. The uniformly magnetised cubes with $d=40$~nm side are considered with $M_{\rm s}=160$~kA/m magnetisation.}
  \label{fig:rods}
\end{figure}

The computational cost of the search for the ground states is greatly reduced when we study [001] magnetisation along one side of the cube. The chains have minimal binding energy for \textcolor{black}{$N\le25$}. Beyond \textcolor{black}{$N=25$}, we observe different behaviour for even ({i.e.}, $N=2k$, $k\in\mathbb{N}$) and uneven number ($N=2k+1$) of particles. For an even number of particles \textcolor{black}{$26\le 2k\le70$}, a ribbon composed of two chains is created, see {Figure~\ref{fig:rods}}(A2). The configuration is antiferromagnetic, {i.e.}, the two chains have magnetisation. At this point, energy increase due to breaking one chain into two is offset by the magnetic interaction of the resulting two touching chains. This configuration is followed by a three-dimensional rod with a square $2\times2$ profile between \textcolor{black}{$72\le 4k\le224$}, {cf.} Figure~\ref{fig:rods}(D2). For an uneven number of particles the chains remain stable until \textcolor{black}{$2k+1\le37$}, see Figure~\ref{fig:rods}(C2). Beyond this point, two touching chain configuration, with $k$ and $k+1$ particles becomes energetically favourable. The reason for this is a penalty of the ultimate particle, having only one neighbour, compared to the double chain composed of an even number of particles. As a result, the assemblies with unequally long chains exhibit higher binding energies and retarded transition to thicker rods. A further example is a transition from a double chain with \textcolor{black}{$4k+2$} particles to a square-profile $2\times2$ rod. This transition is taking place at \textcolor{black}{$4k+2\le86$}. The energy differences between competing states are quite small (less than $1\%$ total energy) and also much smaller than the gain in binding energy due to the additional particle.

For simplicity, from now on, we will only consider rods composed of chains with equal length and analyse the evolution of the profile of the rods. We observe an almost smooth energy drop, see Figure~\ref{fig:energy}. This is reminiscent of behaviour seen in rods composed of magnetic spheres\cite{MESSINA201710, Messina_PhysRevE_2014R}. \textcolor{black}{For $N=342$ magnetic nanocubes, we obtain $2\times3\times57$ and for $N=992$ $4\times4\times62$ rods.}  

\begin{figure}
\includegraphics[width = 8 cm]{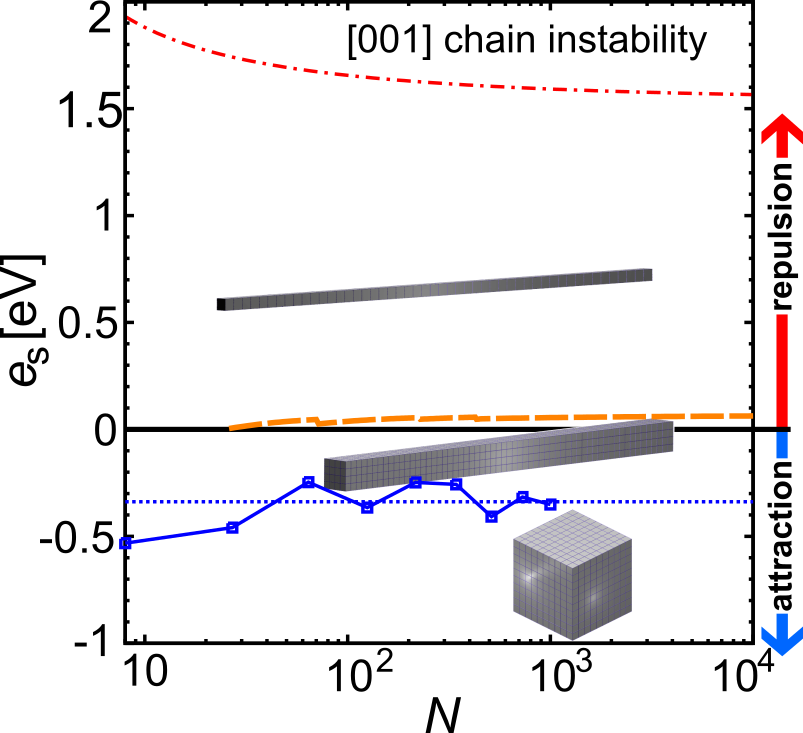}
  \caption{The phase diagram for magnetic [100] nanocubes with a surface coupling $e_{\rm s}$ plotted as a function of the cluster size $N$. The (blue) squares, connected as a guide to the eye, represent the transition from super-rod to regular super-cube for $N=$ 8, 27, 64, 125, 216, 343, 512, 729, and 1000. The dotted blue line is a trend from the energy scaling laws for the energy of super-rods and super-cubes. The dashed (orange) curve is an upper limit of the stability of super-rods above which they transform into head-tail chains. The dash-dotted (red) curve represents the limit of stability of the chains. The uniformly magnetised cubes with $d=40$~nm side are considered to have saturation magnetisation $M_{\rm s}=160$~kA/m. The energies are given in electron-volts, {i.e.}, the reference energy is $\upsilon=1$~eV.}
  \label{fig:diag001}
\end{figure}

\subsubsection{Phase behaviour of [001] magnetised nanocubes: cluster size effects}

Having now a good understanding of the structural evolution of the minimal energy structures composed of [001] magnetised cubes we can capture geometrical properties of the super-rods as a function of $N$.
More specifically, we consider the height $h$ of the super-rod (parallel to the magnetisation axis of the cube) and its profile. Interestingly, while the super-rod growth scales in all three dimensions similarly $N^{1/3}$, the height grows $h\approx\textcolor{black}{6.2}N^{1/3}$ faster than the profile width $w\approx\textcolor{black}{0.4}N^{1/3}$. The binding energy of the rods converges to $u^{\rm 3D, [001]}_{\infty}=-\textcolor{black}{2.15}$~eV, {i.e.}, which is the same energy already found in the super-cubes of [111] magnetised particles. This is not physically surprising since a uniformly magnetised block of material should have the value of the binding energy independent of the orientation of the bulk magnetisation.

In {Figure~\ref{fig:diag001}}, the diagram shows boundaries between different classes of structures build with [001] magnetised nanocubes concerning the number of building blocks and the surface interaction energy. The size effects on the stability of different structures are weak and limited to agglomerates with $N<100$ nanocubes.  The energy of infinite chain is $u^{ht, [001]}_{\infty}=2.01$~eV. Therefore, the limiting surface repulsion energy necessary to prevent creation of very large three dimensional agglomerates is \textcolor{black}{$\epsilon^{[001]}_{\rm S}=(u^{\rm ht, [001]}_{\infty}-u^{\rm 3D}_{\infty})/2\approx0.07$~eV}. 

The surface attraction tends to reduce the external surface and increase the number and size of the contacts between cubes. As we already discussed, super-cube is the most compact super-particle regarding the surface to volume ratio. Based on scaling laws for [001] super-cubes and rods we estimate that the critical surface attraction is given by $\epsilon_{\rm S}\approx\textcolor{black}{-0.4}$~eV for $N\gg1$, beyond which super-cubes will form in [001] systems, {\it  cf.}, Figure~\ref{fig:diag001}. 

\subsection{Surface tension effect: can super-spheres win?}

Up to this point, we have analysed only inter-particle interactions. The suspension can have a profound effect on self-assembly and we will show that magnetic nanocubes are an interesting system in that respect. Using our numerical results we look into magnitudes of surface tension and effective surface area required to induce a transformation from super-cube to super-sphere. Thus effective surface controlled by size-dependent hydration, as predicted by the LCW theory\cite{lum1999}, plays a significant role in the formation of super-particles. Large curvatures, i.e., steps and corners of hydrophobic surfaces, lead to very high local values of surface energy. The suspension liquid cannot fill corners created by kinks and steps of the structure. The exterior surface of nanocubes in agglomerate represents the upper limit for the value of an effective surface and the lower limit is a smooth surface fully enveloping the agglomerate. In our case, the difference between super-cube and super-sphere arises from the geometry of their exterior surface. While the super-sphere has a few, if any, flat surfaces, super-cube has six flat sides. In the case of super-cube, therefore, the effective surface area will be equal to its surface. In the case of super-sphere, there is a range of possible values for the effective surface area depending on the quality of wetting. The upper limit of this range for good wetting is determined by the true surface of the super-sphere, $S^{\rm max}=7.4N^{2/3}d^2$, cf. Table~\ref{table1}. The surface of a perfect sphere with a volume of $Nd^3$ is $S^{\rm min}=(36\pi)^{1/3}N^{2/3}d^2\approx4.83N^{2/3}d^2$ and represents a lower boundary for the effective surface area. To obtain a super-sphere, its effective surface area has to be smaller than the surface of super-cube, i.e, $6N^{2/3}d^2$. In this way, surface energy will off-set the energy penalty for magnetically less-favourable configuration of super-sphere. To have a smaller effective surface of super-sphere, we need a liquid that does not fully wet the surface. 

\begin{figure*}
\includegraphics[width = 17 cm]{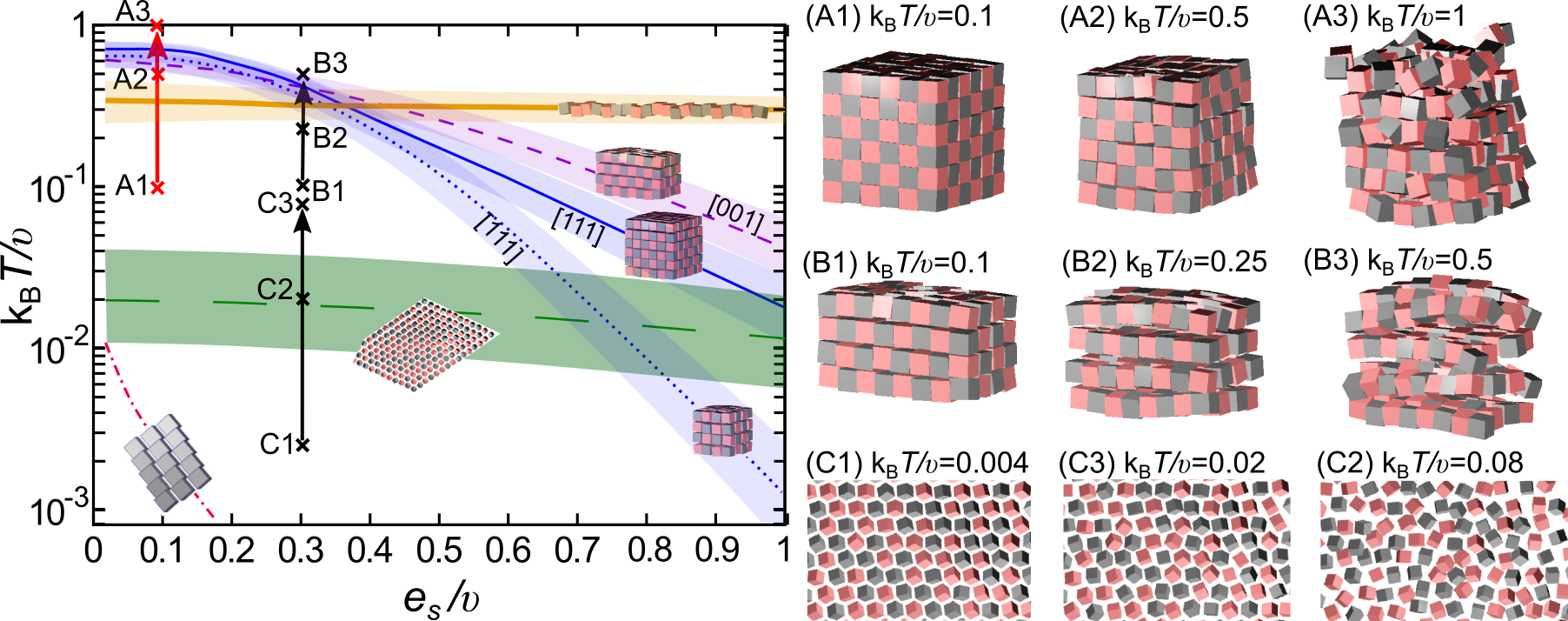}
  \caption{The diagram (left panel) showing the thermal stability region of super-structures made of nanocubes dependence on surface repulsion. The bold (yellow) line shows the stability limit for the [111] chain. The short-dashed (magenta) line shows stability region for [001] magnetised super-rod with $N=128$ nanocubes, dotted (blue) line for [111] magnetised super-cube with $N=64$ nanocubes, full (blue) for [111] magnetised super-cube with $N=216$ nanocubes, and long-dashed line (green) corner-cube flake with $N=180$ nanocubes. The right side panels show snapshots of the evolution of structure with temperature for [111] super-cube (A1, A2, A3), [001] super-rod (B1, B2, B3), and [111] corner-cube two-dimensional structure (C1, C2, C3). The positions of the configurations in parameter space are also shown on the left panel. The results are material and scale-independent. If uniformly magnetised cubes are considered to have saturation magnetisation $M_{\rm s}=160$~kA/m, the reference energy and temperature depend on particle size. For $d=15$ and $40$~nm are $\upsilon=50$~meV and $\upsilon=1$~eV, respectively. The particles are coloured differently to make them more distinguishable.}
  \label{fig:diagT}
\end{figure*}

In the following, we will discuss how the surface tension magnitude $\gamma$ of real liquids and the effective surface area change balance between super-sphere and super-cube. We define the critical surface energy $\gamma_{\rm c}S^{eff}$ as the energy at which super-sphere and super-cube have the same energy. The total energy of the system due to molecular interactions is $U+\gamma_c S^{\rm eff}$ where $U$ is the total interaction energy in the system due to magnetic coupling between particles. We obtain $\gamma_{\rm c} =\frac{\mu_{0}}{4\pi}\frac{\Lambda}{6-\sigma}M_{\rm s}^2d$, where $\sigma$ is the effective surface factor given by $\sigma=[4.83,7.4]$ (the lower boundary stands for no-wetting and the upper one for full wetting of super-sphere surface) and $\Lambda$ is an energy parameter which depends on magnetisation, \textcolor{black}{$\Lambda=0.92$ for [111] and $0.98$} for [001] magnetisation (from scaling laws for energy per particle in Table~\ref{table1}). For a fully spherical effective surface formed by magnetite cubes (magnetisation [001], $M_{\rm s}=160$~kA/m and $d=40$~nm), we obtain $\sigma=4.83$ and $\gamma^{\rm min}_{\rm c}=\textcolor{black}{96}$~\textmu N/m, which is significantly lower than the typical surface energy of any solvent. For smaller magnetic particles the value of $\gamma^{\rm min}_c$ decreases. Therefore, the super-particles wettability predicted by the LCW theory plays a major role in tuning of the equilibrium shape of super-particles by steering the balance between the bulk and the surface energy terms in minimising the overall energy.

\subsection{Finite-temperature behaviour}

Having thoroughly analysed the energies of different configurations at zero temperature limit, we can now proceed to an evaluation of the results obtained from the simulation at finite temperatures. At this point, it is important to note, kinetic energy ${\rm k_B}T$ of the particles enters as a third factor in energy balance that determines the structure along with magnetic and surface energies. If the kinetic energy is comparable to the interaction energies, the thermal motion influences the system. Since the size of the magnetic cubes has a strong effect on their reference magnetic binding energy $\upsilon$, we show results in scaled units $k_{\rm B} T/\upsilon$ in {Figure~\ref{fig:diagT}}. We have followed the evolution of potential energy with temperature for several configurations ({e.g.}, corner-cube flake, [111] super-cube, [001] super-rod, and [111] chain) and define a transition temperature based on the change in heat capacity of the cluster. The delimiting lines and uncertainty regions are interpolated through discrete points obtained from molecular dynamics simulations. We have crosschecked the results of the analysis with snap-shots and there is a clear correlation with the evolution of the structure, {cf.} right panels in Figure~\ref{fig:diagT}. 

At low-temperatures (${\rm k_B}T<<\upsilon$) the associated thermal fluctuations in the configurations are extremely small. We observe the finite-range surface repulsion used in simulation influences the configurations at low temperatures. The super-rod of [001] magnetised cubes shows structural anisotropy, which is originated from the anisotropy of magnetic interaction. The magnetic attraction is two times stronger in magnetisation direction, {i.e.}, the long axis of the super-rod, than in the orthogonal plane. The cubes separate more in directions where the magnetic attraction is weaker, {cf.} Figs.~\ref{fig:diagT}(B1). The cubes are separated by repulsion, while positions and magnetic orientation of the constitutive particles stay unchanged. The dilatation of the structure is a physical consequence of the finite range of the repulsion and in the experiment will depend on how the repulsive interactions are realised in the system, {e.g.}, adsorbed layers, polymers grafted to the surface, etc. We also observe in corner-cube flakes that particles do not touch each other due to repulsion Figs.~\ref{fig:diagT}(C1). With the increase of temperature, the [001] chains of magnetic cubes start to move and particles within chains rotate relatively to each other. With increasing temperature, the space between cubes created by repulsion allows the cubes additional freedom to move, reducing the influence of their shape anisotropy.

At elevated temperatures, in two and three-dimensional structures, both positional and magnetic order disappear while constituent cubes stay connected inside a disordered cluster. Due to stronger binding within the three-dimensional structure for surface repulsion $e_{\rm S}<0.3$, the super-rods are more stable compared to chains, {cf.} dashed curve in Figure~\ref{fig:diagT}. The change of structure with increasing surface repulsion is not isotropic. We observe that with an increasing repulsion, the structure splits into the assembly of aligned head-tail chains (attraction inside the chain is two times stronger than between the chains). The thermal motion makes individual cubes oscillate around their equilibrium positions. As a result, the chains are pushed away from each other to accommodate the movement. At sufficiently elevated temperatures, the bundle of [001] chains is created and lateral order (alignment) disappears. Also, the cubes in these head-tail magnetised chains start to rotate around their magnetisation axis, breaking face-to-face order within the chain. The transition temperature from super-rod to a [001] chain bundle decreases with surface repulsion $e_{\rm S}$.

The distance between single cubes increases homogeneously in [111] super-cube with increasing surface repulsion and temperature. Melting of the cubic structure takes place when the cubes are sufficiently separated and can rotate freely around their magnetic axis. Assembly changes shape and takes the form of a spherical cluster after melting. The melting takes place first at the surface since the surface particles have fewer neighbours to constrain them geometrically and through magnetic interactions. Likewise, we also observe the size stabilisation of the super-cubes, {i.e.}, larger super-cubes melt at a higher temperature. Figure~\ref{fig:diagT} demonstrates size stabilisation. The evolution of the melting temperature with surface energy for $e_{\rm S}$ for $N=64$ and $216$ particle super-cube is denoted with dotted and full (blue) curves, respectively.

For corner-cube flakes, we observe a discrepancy between the stability predicted by analytic theory and simulation. The reason is that simulations produced a far greater separation between edges even for small surface repulsion $e_{\rm S}<0.3$ than seen in the experiment, {cf.} Figure~\ref{fig:diagT}(C1). This resulted in the structures being less stable than the analytic results indicate. The probable reason is that our model does not capture the strong repulsion of the surfaces and weak repulsion between the edges. Our interpretation of the mechanisms of corner structures attributes their formation to the combined strong magnetic attraction between particles in edge contact and surface repulsion, preventing them from closing into a three-dimensional cluster, achievable by grafting polymers on them.

The chain of [111] magnetised cubes stays connected up to ${\rm k}_{\rm B}T/\upsilon<0.35$ even if particles are made very repulsive, i.e., $e_{\rm S}>0.5$ see bold line (yellow) in Figure~\ref{fig:diagT}. The nanocubes to reduce their surface energy by rotating away from full face-to-face contact. In the case of the chain, such rearrangement of particles does not alter long-range order, neither makes chain disconnected.

\section{Discussion}

The competition between shape and magnetic anisotropies offers various pathways for self-assembly. To explore the structure of polymorphs ($N\le27$), analytic calculations are combined with a systematic search for ground states. Two orientations of the dipole relative to the cube geometry were considered, namely in the crystallographic directions [001] and [111]. For magnetic anisotropy in the [001] direction, linear chains have minimal binding energy for $N\le\textcolor{black}{25}$. Beyond  $N=\textcolor{black}{26}$, we observe complex re-entrant behaviour. Between \textcolor{black}{$26\le N\le72$}, for an even number of particles, the ground state is an antiferromagnetic ribbon, and the structure has a closed magnetic flux.  For an uneven number of particles chain remains the ground state and does not enclose the magnetic flux. The strong finite-size effects found in dipolar systems are responsible for the persistence of the chains as the ground state~\cite{MESSINA201710}. Beyond \textcolor{black}{$N\ge72$}, three-dimensional antiferromagnetic rods were found as ground states. In the case of [111] magnetised cubes, the ground state was found to consist of structures derived from a simple square lattice resulting in the super-cubes seen in experiments, {cf.} Figure~\ref{fig:S0}E,G. The surface minimisation and the interplay of the magnetic anisotropy and the cube geometry, which in turn form antiferromagnetic alternating magnetic order, are the driving mechanisms leading to the formation of the super-cube.

The creation of directed non-close-packed arrangements is a technological challenge. We depart in our consideration from ground-state configurations found in polymorphs ($N\le27$) and compare them with the configurations found in experiments. We study the transition between different states with increasing repulsion surface energy. For magnetisation [001], the linear chain becomes minimal energy state for a repulsive surface coupling energies higher than \textcolor{black}{$4\%$} magnetic bulk binding energy. The [111] magnetised cubes will transit into a chain structure only in a range of surface repulsive energies and for small and intermediate number of constitutive particles $N<400$. The limit of stability of compact three-dimensional structures also lies around $60\%$ of the magnetic binding energy of the cubes. The strong repulsion still does not result in the fully stretched chains with cubes touching only at the corners. Instead, the system collapses into a corner-cube two-dimensional configuration with the cubes in contact over the edges. Surface coupling, repulsive or attractive, has a profound influence on the magnetic order of the assembled structures. Dense configurations, such as super-rods or super-cubes, are antiferromagnetic, while non-closed-packed arrangements, such as chains and corner-cube planes, are ferromagnetic.

The results revealed vanishing differences in the binding energy between different super-structures, {cf.}, in Figure~\ref{fig:energy}, due to the insertion of additional nanocubes or between different isomers. Even for the smallest considered assemblies - polymorphs containing only several particles ({cf.},  Figs.~\ref{fig:polymorps} and~\ref{fig:rods}), these differences in energy per particle were comparable to the thermal fluctuations at room temperature. Such a small difference in energy can result in the formation of the non-ground state structures seen in the experimental realisation of polymorphs. The probability of proliferation of non-ground state structures decreases with the magnitude of the energy difference between non-ground states and the ground state. In the case of the super-structures, energy differences due to the addition of a nanocube or between different structures with similar numbers of nanocubes were decreasing with the number of particles following the $N^{-1/3}$ scaling law for the two and three-dimensional structures. Nevertheless, the total energy differences of these structures increase with system size as $N^{2/3}$ leading to the formation of fairly regular super-structures such as super-cubes and rods.

\textcolor{black}{The morphology provides an insight into the particle magnetisation. This is particularly useful in situations where electron holography\cite{Gatel2015} measurements are not possible (e.g., large three dimensional systems or if the system assembly is densely deposited on the substrate). The direction of the net magnetic orientation relative to the cube geometry changes the geometrical structure of the assembly. For chains of nanocubes, the assembly mechanism drives the particles to adopt structures that create a head-tail configuration, very much like chains of magnetic beads\cite{StankovicSM16,StankovicNanoscale}. In the case of [001] direction, this leads to a deep central minimum of magnetic potential energy for the lateral movement of the magnetic particles with respect to their magnetisation and consequently to a quite stiff linear configuration in the presence of surface repulsion \textcolor{black}{($e_{\rm s}>0.07$~eV)} or rod-like structures otherwise. On the contrary, when the magnetisation is along the principal axis, i.e., [111] direction, the structure becomes more flexible under repulsion ($e_{\rm s}\gtrapprox$0.6~\textcolor{black}{eV}) or symmetric super-cubes are formed. We find that the overall picture concerning the magnetization of the 3D nanoparticles assemblies is fairly clear: the individual dipole moments tend to cancel each other in antiferromagnetic configurations. With exception of 2D corner-cube flakes, where we find a ferromagnetic zigzag magnetic configuration.}

\section{Conclusion}

We demonstrate a rich variety of structures and non-close-packed arrangements that can be engineered from nanocubes with a combination of surface interaction and magnetic dipole-dipole coupling. Our results are in good agreement with the experimentally obtained structures. The model studied here approaches systems where interparticle energies of interaction are much higher than the thermal energy of the particles. In most of the structures, particles assemble forming flat surfaces to avoid getting trapped in local minima. We established the relationship between arrangement, [100] and [111] magnetic anisotropy, surface interaction, and size of the system. The results are scalable and our approach is valuable to research aimed at the application of novel magnetic superstructures and to direct the self-assembly of unique structures at different scales. The directed assembly of structures opens up new possibilities in material science in terms of tuning properties through the balance of surface and magnetic interactions, which is of practical relevance in many areas of research, including biomedical materials, energy applications, and complex nanoarchitectures for metamaterial coatings or plasmonic elements.

\section*{Conflicts of interest}
There are no conflicts to declare.

\section*{Acknowledgements}
I. S. acknowledge support of Ministry of Education, Science, and Technological Development of the Republic of Serbia through the Institute of Physics Belgrade. I. S. and C. G. acknowledge the financial support received by ANID PIA/APOYO AFB180002, FONDECYT Grant No. 1201102 and CONICYT MEC80170122. Numerical calculations were run on the PARADOX supercomputing facility at the Scientific Computing Laboratory of the Institute of Physics Belgrade.



\balance


\bibliography{rsc} 

\providecommand*{\mcitethebibliography}{\thebibliography}
\csname @ifundefined\endcsname{endmcitethebibliography}
{\let\endmcitethebibliography\endthebibliography}{}
\begin{mcitethebibliography}{55}
\providecommand*{\natexlab}[1]{#1}
\providecommand*{\mciteSetBstSublistMode}[1]{}
\providecommand*{\mciteSetBstMaxWidthForm}[2]{}
\providecommand*{\mciteBstWouldAddEndPuncttrue}
  {\def\EndOfBibitem{\unskip.}}
\providecommand*{\mciteBstWouldAddEndPunctfalse}
  {\let\EndOfBibitem\relax}
\providecommand*{\mciteSetBstMidEndSepPunct}[3]{}
\providecommand*{\mciteSetBstSublistLabelBeginEnd}[3]{}
\providecommand*{\EndOfBibitem}{}
\mciteSetBstSublistMode{f}
\mciteSetBstMaxWidthForm{subitem}
{(\emph{\alph{mcitesubitemcount}})}
\mciteSetBstSublistLabelBeginEnd{\mcitemaxwidthsubitemform\space}
{\relax}{\relax}

\bibitem[Fern{\'a}ndez-Pacheco \emph{et~al.}(2017)Fern{\'a}ndez-Pacheco,
  Streubel, Fruchart, Hertel, Fischer, and Cowburn]{fernandez2017three}
A.~Fern{\'a}ndez-Pacheco, R.~Streubel, O.~Fruchart, R.~Hertel, P.~Fischer and
  R.~P. Cowburn, \emph{Nature Communications}, 2017, \textbf{8}, 15756\relax
\mciteBstWouldAddEndPuncttrue
\mciteSetBstMidEndSepPunct{\mcitedefaultmidpunct}
{\mcitedefaultendpunct}{\mcitedefaultseppunct}\relax
\EndOfBibitem
\bibitem[Wu \emph{et~al.}(2016)Wu, Mendoza-Garcia, Li, and
  Sun]{overviewassembly1}
L.~Wu, A.~Mendoza-Garcia, Q.~Li and S.~Sun, \emph{Chemical Reviews}, 2016,
  \textbf{116}, 10473--10512\relax
\mciteBstWouldAddEndPuncttrue
\mciteSetBstMidEndSepPunct{\mcitedefaultmidpunct}
{\mcitedefaultendpunct}{\mcitedefaultseppunct}\relax
\EndOfBibitem
\bibitem[Bellido \emph{et~al.}(2012)Bellido, Domingo, Ojea-Jiménez, and
  Ruiz-Molina]{small-nanodev}
E.~Bellido, N.~Domingo, I.~Ojea-Jiménez and D.~Ruiz-Molina, \emph{Small},
  2012, \textbf{8}, 1465--1491\relax
\mciteBstWouldAddEndPuncttrue
\mciteSetBstMidEndSepPunct{\mcitedefaultmidpunct}
{\mcitedefaultendpunct}{\mcitedefaultseppunct}\relax
\EndOfBibitem
\bibitem[Shi and Cheng(2020)]{Shi_adfm.201902301}
Q.~Shi and W.~Cheng, \emph{Advanced Functional Materials}, 2020, \textbf{30},
  1902301\relax
\mciteBstWouldAddEndPuncttrue
\mciteSetBstMidEndSepPunct{\mcitedefaultmidpunct}
{\mcitedefaultendpunct}{\mcitedefaultseppunct}\relax
\EndOfBibitem
\bibitem[Singh \emph{et~al.}(2014)Singh, Chan, Baskin, Gelman, Repnin,
  Kr{\'a}l, and Klajn]{singh2014self}
G.~Singh, H.~Chan, A.~Baskin, E.~Gelman, N.~Repnin, P.~Kr{\'a}l and R.~Klajn,
  \emph{Science}, 2014, \textbf{345}, 1149--1153\relax
\mciteBstWouldAddEndPuncttrue
\mciteSetBstMidEndSepPunct{\mcitedefaultmidpunct}
{\mcitedefaultendpunct}{\mcitedefaultseppunct}\relax
\EndOfBibitem
\bibitem[B\'ecu \emph{et~al.}(2017)B\'ecu, Basler, Kuli{\'{c}}, and
  Kuli{\'{c}}]{kulic}
L.~B\'ecu, M.~Basler, M.~L. Kuli{\'{c}} and I.~M. Kuli{\'{c}}, \emph{The
  European Physical Journal E}, 2017, \textbf{40}, 107\relax
\mciteBstWouldAddEndPuncttrue
\mciteSetBstMidEndSepPunct{\mcitedefaultmidpunct}
{\mcitedefaultendpunct}{\mcitedefaultseppunct}\relax
\EndOfBibitem
\bibitem[Ye \emph{et~al.}(2016)Ye, Pearson, Dolbashian, Pstrak, Mohtasebzadeh,
  Fellows, Mefford, and Crawford]{adfm201504749}
L.~Ye, T.~Pearson, C.~Dolbashian, P.~Pstrak, A.~R. Mohtasebzadeh, B.~Fellows,
  O.~T. Mefford and T.~M. Crawford, \emph{Advanced Functional Materials}, 2016,
  \textbf{26}, 3983--3989\relax
\mciteBstWouldAddEndPuncttrue
\mciteSetBstMidEndSepPunct{\mcitedefaultmidpunct}
{\mcitedefaultendpunct}{\mcitedefaultseppunct}\relax
\EndOfBibitem
\bibitem[Xue \emph{et~al.}(2019)Xue, Yan, and Wang]{Xue_adfm.201807658}
Z.~Xue, C.~Yan and T.~Wang, \emph{Advanced Functional Materials}, 2019,
  \textbf{29}, 1807658\relax
\mciteBstWouldAddEndPuncttrue
\mciteSetBstMidEndSepPunct{\mcitedefaultmidpunct}
{\mcitedefaultendpunct}{\mcitedefaultseppunct}\relax
\EndOfBibitem
\bibitem[Li \emph{et~al.}(2020)Li, Yang, and Yin]{Li.adfm.201903467}
Z.~Li, F.~Yang and Y.~Yin, \emph{Advanced Functional Materials}, 2020,
  \textbf{30}, 1903467\relax
\mciteBstWouldAddEndPuncttrue
\mciteSetBstMidEndSepPunct{\mcitedefaultmidpunct}
{\mcitedefaultendpunct}{\mcitedefaultseppunct}\relax
\EndOfBibitem
\bibitem[Butter \emph{et~al.}(2003)Butter, Bomans, Frederik, Vroege, and
  Philipse]{Butter2003}
K.~Butter, P.~H.~H. Bomans, P.~M. Frederik, G.~J. Vroege and A.~P. Philipse,
  \emph{Nature Materials}, 2003, \textbf{2}, 88--91\relax
\mciteBstWouldAddEndPuncttrue
\mciteSetBstMidEndSepPunct{\mcitedefaultmidpunct}
{\mcitedefaultendpunct}{\mcitedefaultseppunct}\relax
\EndOfBibitem
\bibitem[Wu \emph{et~al.}(2014)Wu, Jubert, Berman, Imaino, Nelson, Zhu, Zhang,
  and Sun]{MagneticRecording}
L.~Wu, P.-O. Jubert, D.~Berman, W.~Imaino, A.~Nelson, H.~Zhu, S.~Zhang and
  S.~Sun, \emph{Nano Letters}, 2014, \textbf{14}, 3395--3399\relax
\mciteBstWouldAddEndPuncttrue
\mciteSetBstMidEndSepPunct{\mcitedefaultmidpunct}
{\mcitedefaultendpunct}{\mcitedefaultseppunct}\relax
\EndOfBibitem
\bibitem[Bian \emph{et~al.}(2018)Bian, Chen, Zheng, Du, Lu, Liu, Hu, and
  Zhang]{small-array}
B.~Bian, G.~Chen, Q.~Zheng, J.~Du, H.~Lu, J.~P. Liu, Y.~Hu and Z.~Zhang,
  \emph{Small}, 2018, \textbf{14}, 1801184\relax
\mciteBstWouldAddEndPuncttrue
\mciteSetBstMidEndSepPunct{\mcitedefaultmidpunct}
{\mcitedefaultendpunct}{\mcitedefaultseppunct}\relax
\EndOfBibitem
\bibitem[Li \emph{et~al.}(2019)Li, Wang, Zhang, Wang, Xu, and
  Yin]{Li2019acsnanolett}
Z.~Li, M.~Wang, X.~Zhang, D.~Wang, W.~Xu and Y.~Yin, \emph{Nano Letters}, 2019,
  \textbf{19}, 6673--6680\relax
\mciteBstWouldAddEndPuncttrue
\mciteSetBstMidEndSepPunct{\mcitedefaultmidpunct}
{\mcitedefaultendpunct}{\mcitedefaultseppunct}\relax
\EndOfBibitem
\bibitem[Rossi \emph{et~al.}(2018)Rossi, Donaldson, Meijer, Petukhov, Kleckner,
  Kantorovich, Irvine, Philipse, and Sacanna]{donaldson2018SM}
L.~Rossi, J.~G. Donaldson, J.-M. Meijer, A.~V. Petukhov, D.~Kleckner, S.~S.
  Kantorovich, W.~T.~M. Irvine, A.~P. Philipse and S.~Sacanna, \emph{Soft
  Matter}, 2018, \textbf{14}, 1080--1087\relax
\mciteBstWouldAddEndPuncttrue
\mciteSetBstMidEndSepPunct{\mcitedefaultmidpunct}
{\mcitedefaultendpunct}{\mcitedefaultseppunct}\relax
\EndOfBibitem
\bibitem[Sacanna \emph{et~al.}(2012)Sacanna, Rossi, and Pine]{Sacanna2012}
S.~Sacanna, L.~Rossi and D.~J. Pine, \emph{Journal of the American Chemical
  Society}, 2012, \textbf{134}, 6112--6115\relax
\mciteBstWouldAddEndPuncttrue
\mciteSetBstMidEndSepPunct{\mcitedefaultmidpunct}
{\mcitedefaultendpunct}{\mcitedefaultseppunct}\relax
\EndOfBibitem
\bibitem[Aoshima \emph{et~al.}(2012)Aoshima, Ozaki, and Satoh]{Aoshima2012}
M.~Aoshima, M.~Ozaki and A.~Satoh, \emph{The Journal of Physical Chemistry C},
  2012, \textbf{116}, 17862--17871\relax
\mciteBstWouldAddEndPuncttrue
\mciteSetBstMidEndSepPunct{\mcitedefaultmidpunct}
{\mcitedefaultendpunct}{\mcitedefaultseppunct}\relax
\EndOfBibitem
\bibitem[Balcells \emph{et~al.}(2019)Balcells, Stankovi\'{c},
  Konstantinovi\'{c}, Alagh, Fuentes, L\'opez-Mir, Or\'o, Mestres, Garc\'ia,
  Pomar, and Mart\'inez]{NanoscaleBCL}
L.~Balcells, I.~Stankovi\'{c}, Z.~Konstantinovi\'{c}, A.~Alagh, V.~Fuentes,
  L.~L\'opez-Mir, J.~Or\'o, N.~Mestres, C.~Garc\'ia, A.~Pomar and
  B.~Mart\'inez, \emph{Nanoscale}, 2019, \textbf{11}, 14194--14202\relax
\mciteBstWouldAddEndPuncttrue
\mciteSetBstMidEndSepPunct{\mcitedefaultmidpunct}
{\mcitedefaultendpunct}{\mcitedefaultseppunct}\relax
\EndOfBibitem
\bibitem[Mehdizadeh~Taheri \emph{et~al.}(2015)Mehdizadeh~Taheri, Michaelis,
  Friedrich, F{\"o}rster, Drechsler, R{\"o}mer, B{\"o}secke, Narayanan, Weber,
  Rehberg, Rosenfeldt, and F{\"o}rster]{Mehdizadeh14484}
S.~Mehdizadeh~Taheri, M.~Michaelis, T.~Friedrich, B.~F{\"o}rster, M.~Drechsler,
  F.~M. R{\"o}mer, P.~B{\"o}secke, T.~Narayanan, B.~Weber, I.~Rehberg,
  S.~Rosenfeldt and S.~F{\"o}rster, \emph{Proceedings of the National Academy
  of Sciences}, 2015, \textbf{112}, 14484--14489\relax
\mciteBstWouldAddEndPuncttrue
\mciteSetBstMidEndSepPunct{\mcitedefaultmidpunct}
{\mcitedefaultendpunct}{\mcitedefaultseppunct}\relax
\EndOfBibitem
\bibitem[Wang \emph{et~al.}(2012)Wang, Wang, LaMontagne, Wang, Wang, and
  Cao]{Wang_ja308962w}
T.~Wang, X.~Wang, D.~LaMontagne, Z.~Wang, Z.~Wang and Y.~C. Cao, \emph{Journal
  of the American Chemical Society}, 2012, \textbf{134}, 18225--18228\relax
\mciteBstWouldAddEndPuncttrue
\mciteSetBstMidEndSepPunct{\mcitedefaultmidpunct}
{\mcitedefaultendpunct}{\mcitedefaultseppunct}\relax
\EndOfBibitem
\bibitem[Cuya~Huaman \emph{et~al.}(2011)Cuya~Huaman, Fukao, Shinoda, and
  Jeyadevan]{Huaman2011}
J.~L. Cuya~Huaman, S.~Fukao, K.~Shinoda and B.~Jeyadevan, \emph{CrystEngComm},
  2011, \textbf{13}, 3364--3369\relax
\mciteBstWouldAddEndPuncttrue
\mciteSetBstMidEndSepPunct{\mcitedefaultmidpunct}
{\mcitedefaultendpunct}{\mcitedefaultseppunct}\relax
\EndOfBibitem
\bibitem[Gao \emph{et~al.}(2012)Gao, Arya, and Tao]{Gao2012}
B.~Gao, G.~Arya and A.~R. Tao, \emph{Nature Nanotechnology}, 2012, \textbf{7},
  433--437\relax
\mciteBstWouldAddEndPuncttrue
\mciteSetBstMidEndSepPunct{\mcitedefaultmidpunct}
{\mcitedefaultendpunct}{\mcitedefaultseppunct}\relax
\EndOfBibitem
\bibitem[Dey \emph{et~al.}(2017)Dey, Chaudhuri, Ghosh, and
  Goswami]{catalyticcubes}
C.~Dey, A.~Chaudhuri, A.~Ghosh and M.~M. Goswami, \emph{ChemCatChem}, 2017,
  \textbf{9}, 1953--1959\relax
\mciteBstWouldAddEndPuncttrue
\mciteSetBstMidEndSepPunct{\mcitedefaultmidpunct}
{\mcitedefaultendpunct}{\mcitedefaultseppunct}\relax
\EndOfBibitem
\bibitem[Disch \emph{et~al.}(2011)Disch, Wetterskog, Hermann, Salazar-Alvarez,
  Busch, Br{\"u}ckel, Bergstr{\"o}m, and Kamali]{Disch2011}
S.~Disch, E.~Wetterskog, R.~P. Hermann, G.~Salazar-Alvarez, P.~Busch,
  T.~Br{\"u}ckel, L.~Bergstr{\"o}m and S.~Kamali, \emph{Nano Letters}, 2011,
  \textbf{11}, 1651--1656\relax
\mciteBstWouldAddEndPuncttrue
\mciteSetBstMidEndSepPunct{\mcitedefaultmidpunct}
{\mcitedefaultendpunct}{\mcitedefaultseppunct}\relax
\EndOfBibitem
\bibitem[Ahniyaz \emph{et~al.}(2007)Ahniyaz, Sakamoto, and
  Bergstr{\"o}m]{Ahniyaz17570}
A.~Ahniyaz, Y.~Sakamoto and L.~Bergstr{\"o}m, \emph{Proceedings of the National
  Academy of Sciences}, 2007, \textbf{104}, 17570--17574\relax
\mciteBstWouldAddEndPuncttrue
\mciteSetBstMidEndSepPunct{\mcitedefaultmidpunct}
{\mcitedefaultendpunct}{\mcitedefaultseppunct}\relax
\EndOfBibitem
\bibitem[Zhang \emph{et~al.}(2007)Zhang, Zhang, and Glotzer]{Zhang2007}
X.~Zhang, Z.~Zhang and S.~C. Glotzer, \emph{The Journal of Physical Chemistry
  C}, 2007, \textbf{111}, 4132--4137\relax
\mciteBstWouldAddEndPuncttrue
\mciteSetBstMidEndSepPunct{\mcitedefaultmidpunct}
{\mcitedefaultendpunct}{\mcitedefaultseppunct}\relax
\EndOfBibitem
\bibitem[John \emph{et~al.}(2004)John, Stroock, and Escobedo]{John2014}
B.~S. John, A.~Stroock and F.~A. Escobedo, \emph{The Journal of Chemical
  Physics}, 2004, \textbf{120}, 9383--9389\relax
\mciteBstWouldAddEndPuncttrue
\mciteSetBstMidEndSepPunct{\mcitedefaultmidpunct}
{\mcitedefaultendpunct}{\mcitedefaultseppunct}\relax
\EndOfBibitem
\bibitem[Satoh(2017)]{satoh2017modeling}
A.~Satoh, \emph{Modeling of magnetic particle suspensions for simulations}, CRC
  Press, 2017\relax
\mciteBstWouldAddEndPuncttrue
\mciteSetBstMidEndSepPunct{\mcitedefaultmidpunct}
{\mcitedefaultendpunct}{\mcitedefaultseppunct}\relax
\EndOfBibitem
\bibitem[Donaldson and Kantorovich(2015)]{donaldson2015directional}
J.~G. Donaldson and S.~S. Kantorovich, \emph{Nanoscale}, 2015, \textbf{7},
  3217--3228\relax
\mciteBstWouldAddEndPuncttrue
\mciteSetBstMidEndSepPunct{\mcitedefaultmidpunct}
{\mcitedefaultendpunct}{\mcitedefaultseppunct}\relax
\EndOfBibitem
\bibitem[Heinz \emph{et~al.}(2017)Heinz, Pramanik, Heinz, Ding, Mishra,
  Marchon, Flatt, Estrela-Lopis, Llop, Moya, and Ziolo]{HEINZ20171}
H.~Heinz, C.~Pramanik, O.~Heinz, Y.~Ding, R.~K. Mishra, D.~Marchon, R.~J.
  Flatt, I.~Estrela-Lopis, J.~Llop, S.~Moya and R.~F. Ziolo, \emph{Surface
  Science Reports}, 2017, \textbf{72}, 1 -- 58\relax
\mciteBstWouldAddEndPuncttrue
\mciteSetBstMidEndSepPunct{\mcitedefaultmidpunct}
{\mcitedefaultendpunct}{\mcitedefaultseppunct}\relax
\EndOfBibitem
\bibitem[Heuer-Jungemann \emph{et~al.}(2019)Heuer-Jungemann, Feliu, Bakaimi,
  Hamaly, Alkilany, Chakraborty, Masood, Casula, Kostopoulou, Oh, Susumu,
  Stewart, Medintz, Stratakis, Parak, and Kanaras]{Heuer2019}
A.~Heuer-Jungemann, N.~Feliu, I.~Bakaimi, M.~Hamaly, A.~Alkilany,
  I.~Chakraborty, A.~Masood, M.~F. Casula, A.~Kostopoulou, E.~Oh, K.~Susumu,
  M.~H. Stewart, I.~L. Medintz, E.~Stratakis, W.~J. Parak and A.~G. Kanaras,
  \emph{Chemical Reviews}, 2019, \textbf{119}, 4819--4880\relax
\mciteBstWouldAddEndPuncttrue
\mciteSetBstMidEndSepPunct{\mcitedefaultmidpunct}
{\mcitedefaultendpunct}{\mcitedefaultseppunct}\relax
\EndOfBibitem
\bibitem[Niculaes \emph{et~al.}(2017)Niculaes, Lak, Anyfantis, Marras, Laslett,
  Avugadda, Cassani, Serantes, Hovorka, Chantrell, and
  Pellegrino]{Niculaes2017}
D.~Niculaes, A.~Lak, G.~C. Anyfantis, S.~Marras, O.~Laslett, S.~K. Avugadda,
  M.~Cassani, D.~Serantes, O.~Hovorka, R.~Chantrell and T.~Pellegrino,
  \emph{ACS Nano}, 2017, \textbf{11}, 12121--12133\relax
\mciteBstWouldAddEndPuncttrue
\mciteSetBstMidEndSepPunct{\mcitedefaultmidpunct}
{\mcitedefaultendpunct}{\mcitedefaultseppunct}\relax
\EndOfBibitem
\bibitem[Bedanta and Kleemann(2008)]{Bedanta2008}
S.~Bedanta and W.~Kleemann, \emph{Journal of Physics D: Applied Physics}, 2008,
  \textbf{42}, 013001\relax
\mciteBstWouldAddEndPuncttrue
\mciteSetBstMidEndSepPunct{\mcitedefaultmidpunct}
{\mcitedefaultendpunct}{\mcitedefaultseppunct}\relax
\EndOfBibitem
\bibitem[Noh \emph{et~al.}(2012)Noh, Na, Jang, Lee, Lee, Moon, Lim, Shin, and
  Cheon]{Noh2012}
S.~Noh, W.~Na, J.~Jang, J.-H. Lee, E.~J. Lee, S.~H. Moon, Y.~Lim, J.-S. Shin
  and J.~Cheon, \emph{Nano Letters}, 2012, \textbf{12}, 3716--3721\relax
\mciteBstWouldAddEndPuncttrue
\mciteSetBstMidEndSepPunct{\mcitedefaultmidpunct}
{\mcitedefaultendpunct}{\mcitedefaultseppunct}\relax
\EndOfBibitem
\bibitem[Kostiainen \emph{et~al.}(2013)Kostiainen, Hiekkataipale, Laiho,
  Lemieux, Seitsonen, Ruokolainen, and Ceci]{Kostiainen2013}
M.~A. Kostiainen, P.~Hiekkataipale, A.~Laiho, V.~Lemieux, J.~Seitsonen,
  J.~Ruokolainen and P.~Ceci, \emph{Nature Nanotechnology}, 2013, \textbf{8},
  52--56\relax
\mciteBstWouldAddEndPuncttrue
\mciteSetBstMidEndSepPunct{\mcitedefaultmidpunct}
{\mcitedefaultendpunct}{\mcitedefaultseppunct}\relax
\EndOfBibitem
\bibitem[Luo \emph{et~al.}(2015)Luo, Yan, and Wang]{Luo_smll.201501783}
D.~Luo, C.~Yan and T.~Wang, \emph{Small}, 2015, \textbf{11}, 5984--6008\relax
\mciteBstWouldAddEndPuncttrue
\mciteSetBstMidEndSepPunct{\mcitedefaultmidpunct}
{\mcitedefaultendpunct}{\mcitedefaultseppunct}\relax
\EndOfBibitem
\bibitem[Song and Zhang(2004)]{Song2004}
Q.~Song and Z.~J. Zhang, \emph{Journal of the American Chemical Society}, 2004,
  \textbf{126}, 6164--6168\relax
\mciteBstWouldAddEndPuncttrue
\mciteSetBstMidEndSepPunct{\mcitedefaultmidpunct}
{\mcitedefaultendpunct}{\mcitedefaultseppunct}\relax
\EndOfBibitem
\bibitem[Abenojar \emph{et~al.}(2018)Abenojar, Wickramasinghe, Ju, Uppaluri,
  Klika, George, Barsoum, Frangiamore, Higuera-Rueda, and
  Samia]{abenojar2018magnetic}
E.~C. Abenojar, S.~Wickramasinghe, M.~Ju, S.~Uppaluri, A.~Klika, J.~George,
  W.~Barsoum, S.~J. Frangiamore, C.~A. Higuera-Rueda and A.~C.~S. Samia,
  \emph{ACS infectious diseases}, 2018, \textbf{4}, 1246--1256\relax
\mciteBstWouldAddEndPuncttrue
\mciteSetBstMidEndSepPunct{\mcitedefaultmidpunct}
{\mcitedefaultendpunct}{\mcitedefaultseppunct}\relax
\EndOfBibitem
\bibitem[Kronast \emph{et~al.}(2011)Kronast, Friedenberger, Ollefs, Gliga,
  Tati-Bismaths, Thies, Ney, Weber, Hassel, Römer, Trunova, Wirtz, Hertel,
  Dürr, and Farle]{Kronast2011}
F.~Kronast, N.~Friedenberger, K.~Ollefs, S.~Gliga, L.~Tati-Bismaths, R.~Thies,
  A.~Ney, R.~Weber, C.~Hassel, F.~M. Römer, A.~V. Trunova, C.~Wirtz,
  R.~Hertel, H.~A. Dürr and M.~Farle, \emph{Nano Letters}, 2011, \textbf{11},
  1710--1715\relax
\mciteBstWouldAddEndPuncttrue
\mciteSetBstMidEndSepPunct{\mcitedefaultmidpunct}
{\mcitedefaultendpunct}{\mcitedefaultseppunct}\relax
\EndOfBibitem
\bibitem[Wu \emph{et~al.}(2016)Wu, Mendoza-Garcia, Li, and Sun]{wu2016organic}
L.~Wu, A.~Mendoza-Garcia, Q.~Li and S.~Sun, \emph{Chemical reviews}, 2016,
  \textbf{116}, 10473--10512\relax
\mciteBstWouldAddEndPuncttrue
\mciteSetBstMidEndSepPunct{\mcitedefaultmidpunct}
{\mcitedefaultendpunct}{\mcitedefaultseppunct}\relax
\EndOfBibitem
\bibitem[Chen \emph{et~al.}(2004)Chen, Liu, and Sun]{Chen2004}
M.~Chen, J.~P. Liu and S.~Sun, \emph{Journal of the American Chemical Society},
  2004, \textbf{126}, 8394--8395\relax
\mciteBstWouldAddEndPuncttrue
\mciteSetBstMidEndSepPunct{\mcitedefaultmidpunct}
{\mcitedefaultendpunct}{\mcitedefaultseppunct}\relax
\EndOfBibitem
\bibitem[Chou \emph{et~al.}(2009)Chou, Zhu, Neeleshwar, Chen, Chen, and
  Chen]{chou2009controlled}
S.-W. Chou, C.-L. Zhu, S.~Neeleshwar, C.-L. Chen, Y.-Y. Chen and C.-C. Chen,
  \emph{Chemistry of materials}, 2009, \textbf{21}, 4955--4961\relax
\mciteBstWouldAddEndPuncttrue
\mciteSetBstMidEndSepPunct{\mcitedefaultmidpunct}
{\mcitedefaultendpunct}{\mcitedefaultseppunct}\relax
\EndOfBibitem
\bibitem[Gatel \emph{et~al.}(2015)Gatel, Bonilla, Meffre, Snoeck,
  Warot-Fonrose, Chaudret, Lacroix, and Blon]{Gatel2015}
C.~Gatel, F.~J. Bonilla, A.~Meffre, E.~Snoeck, B.~Warot-Fonrose, B.~Chaudret,
  L.-M. Lacroix and T.~Blon, \emph{Nano Letters}, 2015, \textbf{15},
  6952--6957\relax
\mciteBstWouldAddEndPuncttrue
\mciteSetBstMidEndSepPunct{\mcitedefaultmidpunct}
{\mcitedefaultendpunct}{\mcitedefaultseppunct}\relax
\EndOfBibitem
\bibitem[Moya \emph{et~al.}(2017)Moya, Abdelgawad, Nambiar, and
  Majetich]{Moya_2017}
C.~Moya, A.~M. Abdelgawad, N.~Nambiar and S.~A. Majetich, \emph{Journal of
  Physics D: Applied Physics}, 2017, \textbf{50}, 325003\relax
\mciteBstWouldAddEndPuncttrue
\mciteSetBstMidEndSepPunct{\mcitedefaultmidpunct}
{\mcitedefaultendpunct}{\mcitedefaultseppunct}\relax
\EndOfBibitem
\bibitem[Håkonsen \emph{et~al.}(2019)Håkonsen, Singh, Normile, De~Toro,
  Wahlström, He, and Zhang]{Hakonsen_adfm201904825}
V.~Håkonsen, G.~Singh, P.~S. Normile, J.~A. De~Toro, E.~Wahlström, J.~He and
  Z.~Zhang, \emph{Advanced Functional Materials}, 2019, \textbf{29},
  1904825\relax
\mciteBstWouldAddEndPuncttrue
\mciteSetBstMidEndSepPunct{\mcitedefaultmidpunct}
{\mcitedefaultendpunct}{\mcitedefaultseppunct}\relax
\EndOfBibitem
\bibitem[Maeda and Maeda(2015)]{Maeda2015}
H.~Maeda and Y.~Maeda, \emph{Langmuir}, 2015, \textbf{31}, 7251--7263\relax
\mciteBstWouldAddEndPuncttrue
\mciteSetBstMidEndSepPunct{\mcitedefaultmidpunct}
{\mcitedefaultendpunct}{\mcitedefaultseppunct}\relax
\EndOfBibitem
\bibitem[Israelachvili(2011)]{ISRAELACHVILI2011253}
J.~N. Israelachvili, \emph{Intermolecular and Surface Forces}, Academic Press,
  Boston, MA, USA, 2011, pp. 253 -- 289\relax
\mciteBstWouldAddEndPuncttrue
\mciteSetBstMidEndSepPunct{\mcitedefaultmidpunct}
{\mcitedefaultendpunct}{\mcitedefaultseppunct}\relax
\EndOfBibitem
\bibitem[Baskin \emph{et~al.}(2012)Baskin, Lo, and Kr{\'a}l]{Baskin2012}
A.~Baskin, W.-Y. Lo and P.~Kr{\'a}l, \emph{ACS Nano}, 2012, \textbf{6},
  6083--6090\relax
\mciteBstWouldAddEndPuncttrue
\mciteSetBstMidEndSepPunct{\mcitedefaultmidpunct}
{\mcitedefaultendpunct}{\mcitedefaultseppunct}\relax
\EndOfBibitem
\bibitem[Lum \emph{et~al.}(1999)Lum, Chandler, and Weeks]{lum1999}
K.~Lum, D.~Chandler and J.~D. Weeks, \emph{The Journal of Physical Chemistry
  B}, 1999, \textbf{103}, 4570--4577\relax
\mciteBstWouldAddEndPuncttrue
\mciteSetBstMidEndSepPunct{\mcitedefaultmidpunct}
{\mcitedefaultendpunct}{\mcitedefaultseppunct}\relax
\EndOfBibitem
\bibitem[Chandler(2005)]{Chandler2005}
D.~Chandler, \emph{Nature}, 2005, \textbf{437}, 640--647\relax
\mciteBstWouldAddEndPuncttrue
\mciteSetBstMidEndSepPunct{\mcitedefaultmidpunct}
{\mcitedefaultendpunct}{\mcitedefaultseppunct}\relax
\EndOfBibitem
\bibitem[Plimpton(1995)]{plimpton1995fast}
S.~Plimpton, \emph{Journal of computational physics}, 1995, \textbf{117},
  1--19\relax
\mciteBstWouldAddEndPuncttrue
\mciteSetBstMidEndSepPunct{\mcitedefaultmidpunct}
{\mcitedefaultendpunct}{\mcitedefaultseppunct}\relax
\EndOfBibitem
\bibitem[Kamberaj \emph{et~al.}(2005)Kamberaj, Low, and Neal]{Kamberaj2005}
H.~Kamberaj, R.~J. Low and M.~P. Neal, \emph{The Journal of Chemical Physics},
  2005, \textbf{122}, 224114\relax
\mciteBstWouldAddEndPuncttrue
\mciteSetBstMidEndSepPunct{\mcitedefaultmidpunct}
{\mcitedefaultendpunct}{\mcitedefaultseppunct}\relax
\EndOfBibitem
\bibitem[Messina and Stankovi\'{c}(2017)]{MESSINA201710}
R.~Messina and I.~Stankovi\'{c}, \emph{Physica A: Statistical Mechanics and its
  Applications}, 2017, \textbf{466}, 10 -- 20\relax
\mciteBstWouldAddEndPuncttrue
\mciteSetBstMidEndSepPunct{\mcitedefaultmidpunct}
{\mcitedefaultendpunct}{\mcitedefaultseppunct}\relax
\EndOfBibitem
\bibitem[Stankovi{\'c} \emph{et~al.}(2016)Stankovi{\'c}, Da{\v{s}}i{\'c}, and
  Messina]{StankovicSM16}
I.~Stankovi{\'c}, M.~Da{\v{s}}i{\'c} and R.~Messina, \emph{Soft matter}, 2016,
  \textbf{12}, 3056--3065\relax
\mciteBstWouldAddEndPuncttrue
\mciteSetBstMidEndSepPunct{\mcitedefaultmidpunct}
{\mcitedefaultendpunct}{\mcitedefaultseppunct}\relax
\EndOfBibitem
\bibitem[Stankovic \emph{et~al.}(2019)Stankovic, Dasic, Otalora, and
  Garc\'{i}a]{StankovicNanoscale}
I.~Stankovic, M.~Dasic, J.~A. Otalora and C.~Garc\'{i}a, \emph{Nanoscale},
  2019, \textbf{11}, 2521--2535\relax
\mciteBstWouldAddEndPuncttrue
\mciteSetBstMidEndSepPunct{\mcitedefaultmidpunct}
{\mcitedefaultendpunct}{\mcitedefaultseppunct}\relax
\EndOfBibitem
\bibitem[Messina and Stankovi{\'c}(2015)]{Messina_PhysRevE_2014R}
R.~Messina and I.~Stankovi{\'c}, \emph{Phys. Rev. E}, 2015, \textbf{91},
  057202\relax
\mciteBstWouldAddEndPuncttrue
\mciteSetBstMidEndSepPunct{\mcitedefaultmidpunct}
{\mcitedefaultendpunct}{\mcitedefaultseppunct}\relax
\EndOfBibitem
\end{mcitethebibliography}
\bibliographystyle{rsc} 

\end{document}